\documentclass[dvips]{imsart}

\RequirePackage[OT1]{fontenc}
\RequirePackage{amsthm,amsmath}
\RequirePackage{natbib}
\RequirePackage[colorlinks,citecolor=blue,urlcolor=blue]{hyperref}

\arxiv{arXiv:0000.0000}

\startlocaldefs
\numberwithin{equation}{section}
\theoremstyle{plain}

\endlocaldefs

\usepackage[margin=0.9in]{geometry}

\usepackage{mathrsfs}
\usepackage{amssymb}
\usepackage{epsfig}
\usepackage{url}
\usepackage{amsmath}
\usepackage{natbib}
\usepackage{amsthm}
\usepackage{booktabs}
\usepackage{url}
\usepackage{alltt}
\usepackage{enumerate}
\usepackage{multirow}
\usepackage{bm}
\usepackage{color}
\usepackage{algpseudocode}
\usepackage{algorithm}

\newcommand{\rx}{\mathtt{x}}

\newcommand{\ry}{\mathrm{y}}
\newcommand{\vx}{\mathbf{x}}
\newcommand{\vy}{\mathbf{y}}
\newcommand{\vz}{\mathbf{z}}

\newcommand{\base}{\mathtt{g}}

\newcommand{\sD}{\mathscr{D}}
\newcommand{\sY}{\mathscr{Y}}

\newcommand{\nmathbf}{\bm}

\def\bfI{\nmathbf I}

\def\bfbeta   {\nmathbf \beta}
\def\bfgamma  {\nmathbf \gamma}

\def\bfmu     {\nmathbf \mu}

\def\bfSigma  {\nmathbf \Sigma}

\def\boldfacefake#1{\kern-4pt
    \hbox{ \mathsurround=0pt
    \hbox to 0.4pt{$#1$\hss}\hbox to 0.4pt{$#1$\hss}\hbox {$#1$}}}

\newcommand{\be}{\begin{eqnarray}}
\newcommand{\ee}{\end{eqnarray}}
\newcommand{\ba}{\begin{eqnarray*}}
\newcommand{\ea}{\end{eqnarray*}}

\newcommand{\reals}{\mbox{\rm I\kern-.20em R}}
\newcommand{\sreals}{\mbox{\small \rm I\kern-.20em R}}

\newtheorem{theorem0}{Theorem}
\newtheorem{lemma0}{Lemma}
\newtheorem{remark0}{Remark}
\newtheorem{fact0}{Fact}
\newtheorem{example0}{Example}
\newtheorem{definition0}{Definition}
\newtheorem{corollary0}{Corollary}
\newtheorem{proposition0}{Proposition}
\newtheorem{algorithmY}{Algorithm}
\newtheorem{conjecture0}{Conjecture}

\begin{document}

\begin{frontmatter}
\title{Robust Classification of \\ High Dimension Low Sample Size Data} 
\runtitle{High Dimensional Robust Classification}

\begin{aug}
\author{\fnms{Necla} \snm{G\"und\"uz}\thanksref{t1,m1}\ead[label=e1]{ngunduz@gazi.edu.tr}},
\author{\fnms{Ernest} \snm{Fokou\'e}\thanksref{m2}\ead[label=e2]{epfeqa@rit.edu}}

\thankstext{t1}{Corresponding author}
\runauthor{G\"und\"uz and Fokou\'e}

\affiliation{Gazi University\thanksmark{m1} and Rochester Institute of Technology\thanksmark{m2}}

\address{\thanksmark{m1}Department of Statistics, Faculty of Science\\
Gazi University, Ankara, Turkey\\
\printead{e1}}

\address{\thanksmark{m2}School of Mathematical Sciences\\
Rochester Institute of Technology\\
98 Lomb Memorial Drive, Rochester, NY 14623, USA\\
\printead{e2}}
\end{aug}

\begin{abstract}
The robustification of pattern recognition techniques has been the subject of intense research
in recent years. Despite the multiplicity of papers on the subject, very few articles have deeply explored the topic of robust classification in the high dimension low sample size context. In this work, we explore and compare the predictive performances of robust classification techniques with a special concentration on robust discriminant analysis and robust PCA applied to a wide variety of large $p$  small $n$ data sets. We also explore the performance of random forest by way of comparing and contrasting the differences single model methods and ensemble methods
in this context. Our work reveals that Random Forest, although not inherently designed to be robust to outliers, substantially outperforms the existing techniques specifically designed to achieve robustness. Indeed, random forest emerges as the best predictively on both real life and simulated data.
\end{abstract}


\end{frontmatter}

\section{Introduction}
\noindent We are given a data set  $\mathscr{D}=\{ (\vx_1,\vy_1), \cdots,(\vx_n, \vy_n )\}$ where $\vx_i \in \mathscr{X} \subset \mathbb{R}^{p \times 1}$  and $\vy_i \in \mathcal{Y}$. We specifically focus on the challenging multicategorical classification scenario involving the so-called high dimensional low sample size (HDLSS) datasets, that is, $n\lll p$ or more precisely $p$ much larger than $n$,  and  $\vy_i \in \mathcal{Y} = \{1, 2, \cdots, G\}$, where $G$ represents the number of groups/classes to which a $p$-tuple $\vx$ from the input space $\mathscr{X}$ may belong. We consider the task of building the best predictively optimal estimator $\widehat{f}(\cdot)$ of the underlying true classifier $f(\cdot)$ of the data.  Throughout this paper, we shall use the
average test error  $\mathtt{AVTE}(\cdot)$, as our measure of predictive performance, namely
\begin{eqnarray}
    \label{eq:avte:1}
    \mathtt{AVTE}(\widehat{f}) =\frac{1}{R} \sum_{r=1}^{R} \left\{ \frac{1}{m} \sum_{i=1}^{m} \ell(\vy_{i}^{(r)}, \widehat{f}_{r}(\vx_i^{(r)}))\right\},
\end{eqnarray}
where  $\widehat{f}_{r}(\cdot)$ is the $r$-th  realization of the estimator $\widehat{f}(\cdot)$ built using the training portion of the split of $\mathscr{D}$
into training set and test set, and $\left(\vx_i^{(r)},\vy_i^{(r)}\right)$ is the $i$-th observation from the test set at the $r$-th random replication of the split of $\mathscr{D}$. Here, we use the ubiquitous zero-one loss function defined by
\begin{eqnarray}
    \label{eq:1:2}
    \ell(\vy_{i}^{(r)}, \widehat{f}_{r}(\vx_i^{(r)})) = 1_{\{\vy_{i}^{(r)} \neq \widehat{f}_{r}(\vx_i^{(r)})\}}
                                           = \left\{\begin{array}{ll}
                                                 1 & \mbox{if $\vy_{i}^{(r)} \neq \widehat{f}_{r}(\vx_i^{(r)})$}\\
                                                 0 & \mbox{otherwise}.
                                                     \end{array}\right.
\end{eqnarray}
The pattern recognition literature is filled with techniques created and developed to solve precisely this problem. Amongst others, logistic regression, discriminant analysis, $k-$nearest neighbors, classification trees,
support vector machine, random forests, boosted trees, relevance vector classifiers  and gaussian process classifiers, just to name a few. Most of the literature in classification deals with data scenarios where the number $n$ of observations/instances is much larger than the dimensionality $p$ of the input space $\mathscr{X}$. As stated earlier, this paper considers data sets of a very special kind, namely the so-called High Dimension Low Sample Size (HDLSS) datasets, also known as large $p$ small $n$ data, since for this type of data, $n\lll p$,
i.e., $n$ is much less than $p$. Data sets of this type are very common these days especially from the
fields of study involving  microarray gene expression used in diagnosing and helping cure diseases such as cancer.
As a matter fact, we consider six such data sets in this paper containing information about various forms of cancer,
namely {\it prostate, lymphoma, lung, colon, leukemia,
brain}. Traditional classification techniques like logistic regression, discriminant analysis and
$k-$nearest neighbors fail miserably on this kind of data, mainly due to the fact that the condition $n\lll p$ leads to illposedness,
and thereby the inability of those methods to even have a solution. In the case of $k-$nearest neighbors
for instance,  the $n\lll p$  condition leads to a severe case of the curse of dimensionality, since the concept of neighbor then
becomes loose and ill-defined when the dimension of the input space $\mathscr{X}$ is far larger than the number of observations available
\cite{Kondo:2012:1}. Several approaches have been proposed to achieve optimal classification in this HDLSS context.
One of the earliest is regularized discriminant analysis (RDA) proposed and extensively developed by \cite{Friedman:1974:1},
recently used by authors like  \cite{Guo:2006:1} in for the classification of microarray gene expression data.
There is a vast literature on regularized discriminant analysis and regularized logistic regression, with a good
number of the contributions dedicated to handling classification problems when $n\lll p$.
It is important to note that it is quite typical to have contamination in the  data whenever the
dimensionality  of the input space gets ever larger. The presence of outliers in the data is
hard enough in low dimensional spaces ($n\ggg p$ with $p$ relatively small), let alone in extremely high dimensional
spaces where one now has to contend with both ill-posedness and outliers. Indeed, these situations
trigger the need for both regularization (to deal with high dimensional ill posedness) and
robustification to circumvent the ill-effect of outliers \cite{VandenBranden:2005:1}.
In the context of $n < p$, there is a relatively large literature on robust discriminant
analysis with many of the contributions based on various approaches to robust estimation of both location and
scatter \cite{Filzmoser:2009:2}, \cite{Filzmoser:2011:1},
\cite{Filzmoser:2013:1}, \cite{Pires:2003:1}, \cite{Pires:2010:1}, \cite{Todorov:2007:1},
\cite{Todorov:2009:1}. Unfortunately, apart from \cite{VandenBranden:2005:1}, \cite{Pires:2010:1},
there has not been much work on robust discrimination when $n$ is much less than $p$. In fact,
we will reveal in our computational section that the traditional robust approach based on Minimum
Covariance Determinant (MCD) estimation of the covariance structure fails miserably in the
HDLSS context. The two approaches presented and explored by \cite{VandenBranden:2005:1} and
\cite{Pires:2010:1} appear to still be in the very early stages of development.
In our experimentations, we noticed that those techniques tend to work when $n/p$ is close to $10^{-2}$, but
they all fail or struggle if the ratio $n/p$ gets smaller.
In this paper, we explore both real life data - mainly microarray gene expression cancer data - and simulated data,
and we reveal patterns exhibited by the average test error as a function $n$, $p$ and
$G$. In the context of simulated data, we also consider the impact of the
contamination rate $\epsilon$, the magnitude $\kappa$ of contamination of the scatter matrix,
and the level $\rho$ of correlation among the predictor variables. Throughout our simulations, we use the
same value for the contamination of the location.  The rest of this paper is organized as follows. In section two, we present a brief summary
of discriminant  analysis with an emphasis on where the need for robustification and regularization arises.
In section three, we discuss some of the most commonly used techniques of robustification, highlighting
some of the limitations and merits of each method. In section four, we present the
computational comparison of the techniques on real life data. In section five, we present a large
simulation study, featuring various choices of the sample size $n$, dimensionality $p$,
number of classes $G$, contamination rate  $\epsilon$ and contamination size $\kappa$ and correlation
among the predictive variables $\rho$.  We also highlight how various scenarios of these choices impact
the average prediction error over $R$ replications. In the section six, we present our conclusion
and discussion along with a brief introduction to our future work dedicated to the regularized
version of robust discriminant analysis in the HDLSS context.

\section{{\it Elements of Discriminant Analysis}}

\noindent Discriminant analysis is arguably one of the oldest and most commonly used approaches
to pattern recognition. The main idea is that given $G$ distinct groups one defines $G$ different
functions  $\delta_{k}(\cdot)$, for $k=1,\cdots, G$, such that given a data matrix
${\bf X} = \big[\vx_1^\top,\vx_2^\top,\cdots,\vx_n^\top\big]$ whose $i$th row $\vx_i^\top = (\rx_{i1},
\rx_{i2},\cdots,\rx_{ip})$ for which class label $\vy_i$ is desired,
\begin{eqnarray}
    \label{eq:2:1}
    \widehat{\vy}_i & = & \widehat{class(\vx_i)} = \operatorname*{arg\,max}_{k=1,\cdots, G} \{\widehat{\delta}_{k}(\vx_i)\}
\end{eqnarray}
where $\widehat{\delta}_{k}(\vx_i)$ is the estimator of $\delta_{k}(\vx_i)$. In other words, discriminant
analysis estimates the class of a new object as the label whose discriminant functions yields
the largest value given $\vx_i$. Now, when the density function of $\vx_i$ given class $k$ is  known to be multivariate Gaussian,
we have quadratic discrimination. If in addition all the classes have the same covariance matrix, i.e.
$\left(X \vert k\right)\sim \mathtt{MVN} \left(\bfmu_{k}, \Sigma \right)$,
we are in the presence of linear discriminant analysis (LDA). In other words,
with LDA,  each sample group has its class mean $\mu_{k}$ $\left( k=1,2,\cdots,G \right)$
but shares the sample covariance matrix $\Sigma$ with the other groups. If $\pi_k = \Pr[Y=k]$
denotes the prior probability of class membership, then the LDA discriminant
function is given by

\begin{eqnarray}
    \label{eq:2:3}
    \delta_{k}(\vx) = {-}\frac{1}{2} \left(\vx - \bfmu_{k} \right)^\top \Sigma^{-1} \left(\vx - \bfmu_{k} \right) + \log \pi_{k}.
\end{eqnarray}
In practice,  $\delta_{k}(\cdot)$ is estimated from the data $(\vx_{1},\vy_{1})$, $\cdots$, $(\vx_{n},\vy_{n})$ by its empirical counterpart

\begin{eqnarray}
 \label{eq:2:3}
 \widehat{\delta_{k}(\vx)} = {-}\frac{1}{2} \left(\vx - \widehat{\mu}_{k} \right)^\top {\widehat{\Sigma}}^{-1} \left(\vx - \widehat{\mu}_{k} \right) + \log \widehat{\pi}_{k},
\end{eqnarray}

\noindent where  ${\widehat{\pi}}_{k}=\frac{n_{k}}{n}=\frac{\sum\limits_{i=1}^{n}{w_{ik}}}{n}$ is the observed proportion of group/class $k$ observations and $\widehat{\bfmu}_{k}=\bar{\vx}_{k}=\frac{1}{n_{k}} \sum\limits_{i=1}^{n} w_{ik} \vx_{i}$ is the empirical (sample) mean in class $k$, both defined using $n_{k}=\sum\limits_{i=1}^{n‎} w_{ik}$, the number of observations from class $k$ and  the observed indicator of class membership
given by
\begin{eqnarray}
      \label{eq:2:4}
      w_{ik}=1_{\{\vy_i=k\}} = \left\{\begin{array}{ll}
                           1 & \mbox{if $\vy_i = k$},\\
                           0 & \mbox{otherwise}.
                    \end{array}
             \right.
\end{eqnarray}
Finally, the estimate $\widehat{\Sigma}$ of the common covariance $\Sigma$ is often the pooled covariance given by
\begin{eqnarray}
    \label{eq:2:5}
    \widehat{\Sigma} = \frac{\sum\limits_{k=1}^{G‎} (n_{k}-1) S_{k}}{\sum\limits_{k=1}^{G‎}n_{k}-G},
\end{eqnarray}
\noindent where $S_{k}$ is the sample (empirical) variance-covariance matrix of the $k$th class, namely
\begin{eqnarray}
   \label{eq:2:6}
    S_{k} =\frac{1}{n_{k}-1} \sum\limits_{i=1}^{n‎} w_{ik} \left(\vx - \widehat{\mu}_{k} \right)^\top \left(\vx - \widehat{\mu}_{k} \right).
\end{eqnarray}


\noindent It turns out that both the estimated location $\widehat{\bfmu}_k$ and the estimated scatter matrix
$\widehat{\Sigma}$ are sensitive to outliers, making the estimated discriminant function non robust under contamination.
In the context of prediction, non-robustness leads to poor predictive performances due the fact the presence of outliers in the explanatory variables causes the vital components of the discriminant function to be biased.
We therefore need robust methods to estimate both location and scatter parameters in order to decrease the prediction error.
When in addition we have $p$ much larger than $n$, we encounter the extra problem of
non-invertibility of $\widehat{\Sigma}$. Typically, one can solve this problem
by selecting few subset of variables. However, a more general approach deals with the problem
by regularizing $\widehat{\Sigma}$ using
$\widetilde{\Sigma}=\widehat{\Sigma}+\lambda I_p$ with $\lambda \in (0,\infty)$, or a convex
and more general version of it where
\begin{eqnarray}
\widetilde{\Sigma}=(1-\alpha)\widehat{\Sigma}+ \frac{\alpha}{p}\mathtt{trace}(\widehat{\Sigma}) I_p
\end{eqnarray}
with $\alpha \in (0,1)$. Some of the earliest work on regularized discriminant analysis include the seminar paper by \cite{Friedman:1974:1}, and later applications by \cite{Guo:2006:1}, just to name a few. Many authors have contributed extensively in the area of regularized discriminant analysis.
Clearly, in the presence of both outliers and high dimensionality, one needs to both robustify and then
regularize. In the subsequent section, we mainly focus on the kind of robustification that deals with
contaminated high dimensionality through a suitable combination robust PCA and projection pursuit \cite{Pires:2010:1}, \cite{VandenBranden:2005:1}.

\section{{\it Robust Estimation Methods for Linear Discriminant Analysis}}
\noindent In section two, one the challenges of discriminant analysis came from the fact that in the
presence of outliers, the location and scatter estimates are not
reliable because they are not robust. Many authors have contributed a wide variety of approaches all aimed at
 at robustifying LDA. One of the earliest approaches used to address this problem is a technique known as Minimum Covariance Determinant (MCD) introduced and developed by \cite{Rousseeuw:1984:1}, \cite{Rousseeuw:1985:1}.
The literature on robust discriminant analysis has blossomed recently,
with many papers studying  and exploring various extensions of MCD. In this paper,
we will be comparing the predictive performances of extensions developed
by (\cite{Croux:2001:1}), then the one explored by \cite{He:2000:1} and \cite{HubertVanDriessen:2004:1}.
We will also look into the predictive performances of MCD extensions proposed by (\cite{HubertVanDriessen:2004:1}), and the
 ones developed by \cite{Todorov:2007:1}, \cite{Hawkins:1997:1}. Another paper extending the work on MCD is
  contributed by \cite{HubertDebruyne:2010:1}. It is important to note that all the above mentioned variations on the MCD theme have been implemented in R with packages like {\tt rrcov} and {\tt rrcovHD} readily
available for immediate installation and use. Our goal in this paper is to consider different
scenarios of contamination in high dimensional spaces, and then compare the performances of existing
techniques, with the finality of establishing the conditions under which each one of the techniques performs best.
\cite{Rousseeuw:1984:1}'s original MCD estimator is quite intuitive and can be briefly described as follows:
given $n$ observations $\vx_1, \vx_2, \cdots, \vx_n$ taken from a $p$-dimensional space $\mathscr{X} \subset \mathbb{R}^p$,
with true location vector $\bfmu$ and true scatter matrix $\bfSigma$,
find the subset of $h$ observations out of $n$ such that the corresponding sample (estimated) covariance matrix yields the smallest determinant.
In other words, we must have
$$
{\tt det}(\widehat{\bfSigma}(\bfgamma^{\tt (MCD)})) = \underset{\bfgamma \in \{0,1\}^n}{\min}\left\{{\tt det}(\widehat{\bfSigma}(\bfgamma))\right\}
$$
where $\bfgamma \in \{0,1\}^n$ simply represents the indicator vector such that $\gamma_i = 1$ if observation $i$ is among the $h$ chosen, and
$\gamma_i = 0$ if it is not. Obviously, the final indicator vector $\bfgamma^{\tt (MCD)}$ chosen by the MCD method is such that
${\tt length}(\bfgamma^{\tt (MCD)})=|\bfgamma^{\tt (MCD)}|=h$. Also, we use the notation $\widehat{\theta}(\bfgamma)$ to denote the estimator of $\theta$ based on only the observations selected by the indicator vector $\bfgamma$. Essentially, the MCD estimator of the location parameter $\bfmu$ is defined by the mean of that subset $\bfgamma^{\tt (MCD)}$ and the MCD estimator of the scatter matrix parameter $\bfSigma$ is defined by the covariance of that subset $\bfgamma^{\tt (MCD)}$. More specifically,
$$
\widehat{\bfmu}_{\tt MCD} = \widehat{\bfmu}(\bfgamma^{\tt (MCD)}) \quad \text{and} \quad \widehat{\bfSigma}_{\tt MCD} = \widehat{\bfSigma}(\bfgamma^{\tt (MCD)}).
$$
In practice, $\widehat{\bfSigma}_{\tt MCD}$ is chosen in such a way that it is a multiple of its covariance matrix.
The multiplicative factor is chosen in such a way that $\widehat{\bfSigma}_{\tt MCD}$ is consistent at the multivariate normal model and unbiased for
small samples (\cite{Pison:2002:1}).
Now, the MCD approach requires that $\frac{n}{2} \leq h < n$, with  $h=[(n+p+1)/2]$ yielding the maximal breakdown point.
Central to MCD is the fact that $h$ must be determined or set.
In fact, it should be noted that {\it the MCD estimator cannot be computed when $p>h$, since such a scenario would mean
having a singular covariance matrix for any $h-$subset. It turns out that the whole MCD machinery
needs $n \geq 2 p$ in order to function at all. For our high dimension low sample size problems for which
$p \ggg n$, it is obvious that the basic formulation of MCD does not work.} We discuss a little later
the extension of MCD known as regularized MCD whereby the estimates of interest are obtained
in their regularized version.
Even within the satisfaction of the $n > 2p$ requirement, MCD is essentially very computationally intensive
for the simple reason that the need to select a subset combinatorially to optimize a criterion requires a number
of computing operations that can explode for even small sample sizes.
Many faster versions of the basic MCD algorithm have been suggested, led by \cite{Rousseeuw98afast}.
As we shall see later in our computations, different variants of MCD lead to sometimes drastically
different predictive performances.
In the context of discriminant analysis, one of the obvious limitations of MCD lies in the fact that
it trims all the classes/groups equally. Such an equal treatment of all the groups is potentially inefficient
in the presence of uncontaminated groups. Many other drawbacks of the basic MCD approach to discriminant analysis
have been scrutinized and addressed by authors such as \cite{He:2000:1},
\cite{HubertVanDriessen:2004:1}, \cite{Christmann:2013:1}, \cite{Croux:2001:1}.
Each of these variants was proposed in order to address/solve a perceived
limitation/drawback of the basic MCD approach. Unfortunately, in turns out all these
variants of MCD only work when $p$ is less than $n$. In fact most of them require $n$ to be
at least greater than $2p$. In order to deal with $n$ less than $p$ situations
one had to abandon MCD in its present form. As a matter of fact, we are currently
working on an extension of the MCD approach that combines robustification and regularization to
address HDLSS situation with  $n/p$  arbitrarily very small.
In this paper however,  we explore two non-MCD based techniques, namely robust SIMCA (described later) and projection pursuit
(PP) discriminant analysis. As we will see in the computational section, PP will proved to be quite flexible
but unfortunately fail when $n/p$ gets to small (less than to $10^{-2}$). We will see later
that PP yields relatively poor predictive performances when the number of classes is greater than $2$ and/or
the intrinsic dimensionality of the input space is inherently high.\\

{\it SIMCA approach to High Dimensional Robust Classification: } Soft Independent Modelling of Class Analogies (SIMCA)
was introduced by \cite{Wold:1976:1}. \cite{Bicciato:2003:1} explain that SIMCA ability to classify high dimensional data comes from
the fact that it is based on a clever adaptation of principal component analysis (PCA).
Thanks to the supervised nature of discriminant analysis (class membership known), SIMCA
proceeds by performing principal component analysis in each of the $G$ classes separately.
Essentially, SIMCA can be summarized as an approach that combines robust PCA within each group
based on robust covariance estimation to achieve good predictive performances in classification.
More details on SIMCA can be found in \cite{HubertVanDriessen:2004:1}, \cite{Hubert:2004:1}
\cite{Bicciato:2003:1} and \cite{VandenBranden:2005:1}. SIMCA has been widely applied to areas as diverse as
image analysis, microarray gene expression classification, and many other
fields where data exists with $n$ much less than $p$. An implementation of Robust SIMCA is
provided through the R package {\tt rrcovHD}, and will be used in our comparison
of predictive performances of high dimensional robust classifiers. \\

\noindent {\it Projection Pursuit Approach for Robust Linear Discriminant Analysis: }
From Wikipedia, {\it Projection Pursuit (PP) is a type of statistical technique which involves finding the most "interesting" possible projections in multidimensional data. Often, projections which deviate more from a normal distribution are considered to be more interesting. As each projection is found, the data are reduced by removing the component along that projection, and the process is repeated to find new projections; this is the "pursuit" aspect that motivated the technique known as matching pursuit.
The idea of PP is to locate the projection or projections from high-dimensional space to low-dimensional space that reveal the most details about the structure of the data set. Once an interesting set of projections has been found, existing structures (clusters, surfaces, etc.) can be extracted and analyzed separately.
PP has been widely use for blind source separation, so it is very important in independent component analysis. PP seek one projection at a time such that the extracted signal is as non-Gaussian as possible.}
From the seminal article \cite{Fried74}, many authors like \cite{Fried81}, \cite{Fried87} have developed applications and extensions of PP to wide variety of statistical problems. There have also been many theoretical justifications and discussions of the strengths
and appeal of PP \cite{Huber}, \cite{Hall}. PP has also been used in discriminant analysis and
outlier detection\cite{Pan} and \cite{Polzehl93projectionpursuit}.
\cite{Pires:2010:1} presents one of the most recent developments on PP for robust discriminant analysis in
high dimensional spaces. Her work builds first on the original PP idea
as presented and developed by \cite{Fried74} and \cite{Huber}, and combines robustness strategies and the foundational idea
of PP presented in the above definition, to achieve robust linear discriminant analysis in high dimensional spaces.
Four of the seven techniques compared in this paper are based on the PP approach to robustness linear discriminant analysis in high dimension spaces. As we'll see later though, it will turn out that PP techniques will fail - somewhat catastrophically at times -
when the intrinsic dimensionality of the data is high. This is unsurprising, since the very idea of PP presupposes the existence
of a lower dimensional space as the true/intrinsic basis of the data. \\

{\it Ensemble Learning Approach to Robust Classification: }
It is often common in massive data that selecting a single model does not
lead the optimal prediction. For instance, in the presence of multicollinearity which is almost inevitable
when $p$ is very large, function estimators are typically unstable, as they tend to exhibit rather estimation large variances.
This issue of inflated
variance gets even more amplified in the presence of data contamination. It makes sense that when data contains outliers,
they will cause the variance of estimators to increase.
The now popular {\it bootstrap aggregating} also referred to as {\it bagging} offers one way to
reduce the variance of the estimator by creating an aggregation of bootstrapped versions
of the base estimator. This is an example of ensemble learning, with the aggregation/combination
formed from equally weighted base learners.
Consider regression for instance and let $\widehat{\tt g}^{\tt (b)}(\cdot)$ be the $b$th bootstrap
replication of the estimated base regression function $\widehat{\tt g}(\cdot)$. Then the {\it bagged}
version $\widehat{{f}}^{\tt (bagging)}(\cdot)$ of the estimator $\widehat{f}(\cdot)$ is given by
$$
\widehat{{f}}^{\tt (bagging)}(\vx) = \frac{1}{B}\sum_{b=1}^{B}{\widehat{\tt g}^{\tt (b)}(\vx)}.
$$
If the base learner is a multiple linear regression model estimator $\widehat{\base}(\vx)=\widehat{\beta}_0 + \vx^{\top}{\widehat{\bfbeta}}$, then the $b$th bootstrapped
replicate is  $\widehat{\tt g}^{\tt (b)}(\vx)=\widehat{\beta}_0^{(b)} + \vx^{\top}{\widehat{\bfbeta}^{(b)}}$, and the bagged version
 $\widehat{{f}}^{\tt (bagging)}(\cdot)$ of the estimator $\widehat{f}(\cdot)$
is
$$
\widehat{f}^{\tt (bagging)}(\vx) = \frac{1}{B}\sum_{b=1}^{B}{\left(\widehat{\beta}_0^{(b)} + \vx^{\top}{\widehat{\bfbeta}^{(b)}}\right)}.
$$
In this paper, our focus is on classification in high dimensional space where $p$ is larger than $n$ and
the data also contains some outliers. Now, let's consider a multi-class classification task as defined much earlier
with labels $\ry$ coming from $\mathscr{Y}=\{1,2,\cdots,G\}$ and predictor variables
 $\vx=(\rx_1,\rx_2,\cdots,\rx_p)^\top$  coming from a $p$-dimensional space $\mathscr{X}$.
 Let $\widehat{\tt g}^{\tt (b)}(\cdot)$ be the $b$th bootstrap replication of the estimated base classifier $\widehat{\tt g}(\cdot)$,
such that $(\widehat{\ry})^{(b)} = \widehat{\tt g}^{\tt (b)}(\vx)$ is the $b$th bootstrap  estimated class of $\vx$. The estimated response by bagging is obtained using the majority vote rule, which means the most frequent label throughout the $B$ bootstrap replications.
Namely,
$$
\widehat{f}^{\tt (bagging)}(\vx) = \texttt{Most frequent label in } \widehat{\tt C}^{\tt (B)}(\vx),
$$
where
$$
\widehat{\tt C}^{\tt (B)}(\vx) = \bigg\{\widehat{\tt g}^{\tt (1)}(\vx), \widehat{\tt g}^{\tt (2)}(\vx), \cdots, \widehat{\tt g}^{\tt (B)}(\vx)\bigg\}.
$$
Succinctly, we can write the bagged estimated label of $\vx$ as
\begin{eqnarray*}
\widehat{f}^{\tt (bagging)}(\vx) = {\tt arg} \,\underset{\ry \in \mathscr{Y}}{\tt max}\left\{{\tt freq}_{\widehat{\tt C}^{\tt (B)}(\vx)}(\ry)\right\}
= {\tt arg} \,\underset{\ry \in \mathscr{Y}}{\tt max}\left\{\sum_{b=1}^{B}{\left({\bf 1}_{\{\ry=\widehat{\tt g}^{\tt (b)}(\vx)\}}\right)}\right\}.
\end{eqnarray*}

It must be emphasized that in general, ensembles do not assign equal weights to base learners in the aggregation.\footnote{The general formulation
in the context of regression for instance is
$\widehat{f}^{\tt (bagging)}(\vx) =\sum_{b=1}^{B}{\alpha^{\tt (b)}\widehat{\tt g}^{\tt (b)}(\vx)},$
where the aggregation is often convex, i.e.  $\sum_{b=1}^B{\alpha^{\tt (b)}}=1$.} For our purposes in this paper, it is crucial
to address the issue of how bagging predictors can help achieve robust classification in high dimensional spaces. It turns out that
by combining bagging with subsetting of the input space, we end up with the so-called \textit{Random Subspace Learning (Attribute Bagging)}
which, as we'll argue later, yields very good predictive performances on high dimensional data contaminated with outliers.
Consider the training set $\sD = \{\vz_i=(\vx_i^\top,\ry_i)^\top,\,\,i=1,\cdots,n\}$, where $\vx_i^\top = (\rx_{i1},\cdots,\rx_{ip}) $ and  $\ry_i\in \sY$ are realizations of two random variables $X$ and $Y$ respectively.
Suppose our goal is to use a total of $B$ base learners to build an ensemble estimator $\widehat{f}^{\tt bagging}(\cdot)$ of the underlying function $f(\cdot)$.
Random Subspace Learning (Attribute Bagging) proceeds very much like bagging, with the
   added crucial step consisting of selecting a subset of the variables from the input space for training rather than
   building each base learners using all the $p$ original variables.
\begin{itemize}
   \item {\tt Randomly draw the number $d<p$ of variables to consider}
   \item {\tt Draw without replacement the indices of $d$ variables} 
   \item {\tt Build the $d$-dimensional model}
\end{itemize}
This step is the main ingredient for variable importance estimation and also has the benefit of
circumventing the limitation of bagging due to correlatedness of the trees making up the ensemble.

\begin{algorithm}
\caption{Random Subspace Learning for Model Aggregation}\label{algo:rf:1}
\begin{algorithmic}[1]
\Procedure{RandomForest}{$B$}\Comment{The Random Forest Algorithm for $B$ trees}
\State {\tt Choose a base learner $\widehat{\base}(\cdot)$} \Comment{e.g.:  Trees}
\State {\tt Choose an estimation method} \Comment{e.g.:  Recursive Partitioning}
\For{$b=1$ to $B$}
\State {\tt Draw with replacement} from $\sD$ a bootstrap sample $\sD^{(b)} = \{\vz_1^{(b)},\cdots,\vz_n^{(b)}\}$
\State {\tt Draw without replacement} from $\{1,2,\cdots,p\}$ a subset $\{j_1^{(b)},\cdots,j_d^{(b)}\}$ of $d$ variables
\State {\tt Drop unselected variables} from $\sD^{(b)}$ so that $\sD_{\tt sub}^{(b)}$ is $d$ dimensional
\State {\tt Build the $b$th base learner} $\widehat{\base}^{(b)}(\cdot)$ based on $\sD_{\tt sub}^{(b)}$
\EndFor \label{algo:rf:1}
\State Use the ensemble $\Big\{\widehat{\base}^{(b)}(\cdot),\, b=1,\cdots, B\Big\}$ to form the estimator
\begin{eqnarray*}
\widehat{f}^{\tt (RF)}(\vx) = {\tt arg} \,\underset{\ry \in \mathscr{Y}}{\tt max}\left\{{\tt freq}_{\widehat{\tt C}^{\tt (B)}(\vx)}(\ry)\right\}
= {\tt arg} \,\underset{\ry \in \mathscr{Y}}{\tt max}\left\{\sum_{b=1}^{B}{\left({\bf 1}_{\{\ry=\widehat{\tt g}^{\tt (b)}(\vx)\}}\right)}\right\}.
\end{eqnarray*}
\EndProcedure
\end{algorithmic}
\end{algorithm}

Let $\vz_i \in \sD$ be a random pair in the original training set $\sD$ of size $n$, and consider the
bootstrapped sample  $\sD^{(b)}$ of size $n$ also, generated by sampling with replacement from $\sD$.
Let $\Pr[\vz_i \in \sD^{(b)}]$ represent the proportion of observations from $\sD$ present in $\sD^{(b)}$.
Then it can be shown that
$\Pr[\vz_i \in \sD^{(b)}]
                         = 1- \left(1-\frac{1}{n}\right)^n$
As a result, if $\Pr[\vz_i \notin \sD^{(b)}]=\Pr[O_n]$ denotes the \textit{proportion of observations from $\sD$ not present in $\sD^{(b)}$}, then
\begin{eqnarray}
\label{eq:oob:1}
\Pr[\vz_i \notin \sD^{(b)}]=\left(1-\frac{1}{n}\right)^n = \Pr[O_n].
\end{eqnarray}
It turns out that $\Pr[O_n] \rightarrow e^{-1} \approx 0.37$ as $n \rightarrow \infty$. Which means that roughly about one third
of the training set is not used when building the $b$th bootstrapped based learner. In the context of data contaminated with outliers,
each observation - including outlier - has an asymptotic probability of $e^{-1}$ of not affecting the base learner at hand. It is therefore
reasonable to deduce that by averaging this exclusion of outliers over many replications ($B$ base learners, with $B$ typically large), one
can achieve a robust classifier through bootstrap aggregation. Our conjecture in this regard
is that through its bootstrapping mechanism, each outlier has a probability of  $e^{-1}$ of not affecting the base learner.

\section{Computational Comparison of Predictive Performances}
We now consider comparing the predictive performances of the techniques described earlier
on real life data. Among other things, we present
the apparent (training) error and the true (test) error which in this case is more precisely
average test error over $R$ replications as defined in \eqref{eq:avte:1}.
$$
\mathtt{AVTE}(\widehat{f}) =\frac{1}{R} \sum_{r=1}^{R} \left\{ \frac{1}{m} \sum_{i=1}^{m} \ell (\vy_{i}^{(r)}, \widehat{f}_{r}(\vx_i^{(r)}))\right\}.
$$
Throughout this paper, each replication randomly assigns $2/3$ of the data to the training set and $1/3$ to the test set.
We do not consider a validation set because none of the techniques is based on a tuning parameter. We use $R=200$ replications.
We analyze 7 different datasets, six of which are high dimension low sample size (HDLSS)
microarray gene expression datasets. For clarity and completeness, we present both the tables and the plots depicting
the predictive performances of the methods explored and analyzed.

\subsection{Description of the datasets}
\noindent {\it (i)} {\tt Diabetes data:} Our first data set deals with diabetes. It contains $145$ observations, 3 variables and
three classes. This is obviously not a high dimensional dataset, but we use it here to reveal
the stark differences in performance between methods when one switches from large $n$ small $p$ to large $p$
small $n$. This dataset is available in the R package called {\tt mclust} contributed by \cite{Reaven:1979:1}.
{\it (ii)} {\tt Ceramic pottery data: } This pottery data set was analysed by \cite{SternDescoeudres:1977:1}. Other authors
have used it to test the robustness of their methods, namely \cite{Cooper:1983:1} and later \cite{Pires:2010:1}.
Please note that these first two datasets are qualitatively different from all the other datasets explored in this paper.
Indeed, all the other $5$ remaining datasets we explore here have in common the fact that they are
 all microarray gene expression data sets.
{\it (iii)} {\tt Prostate cancer data:} This data set comes from microarray gene expression profiles/levels on prostate cancer, and is
a subset of a much larger data set from a study by \cite{Stephenson:2005:1}. This dataset has 37 samples classified as recurrent
and 42 as non-recurrent primary prostate tumor.
{\it (iv)} {\tt Lymphoma data:} The following data set deals with Lymphoma and contains $180$
observations and $661$ variables.
{\it (v)} {\tt Lung cancer data:} This dataset on lung cancer ıs just one of many existing lung cancer datasets. The version we explore in this paper
contains $197$ observations and $1000$ variables.
{\it (vi)} {\tt Colon cancer data:} From \cite{Alon:1999:1}, {it contains 62 observations on subjects classified
into two groups (G1: subjects with colon cancer, with 40 observations; G2: healthy subjects, with
22 observations) and measured on 2000 variables (gene expression levels). The aim is to predict, as
accurately as possible, the disease status from the gene expression levels. This is a well known data
set in the modern classification literature (e.g., References from the paper) and the original version
is available in the colonCA R package from Bioconductor. The raw data is not normalized/preprocessed,
which may lead to very bad classification results. Therefore a simple normalization procedure was
applied: the data were log-transformed and after that each row was individually centered using its
median.}
{\it (vii)} {\tt Leukemia data: } In this Leukemia data set, there are $3571$
variables(features), $72$ samples (\cite{Golub:1999:1}).
{\it (viii)} {\tt Brain cancer data: } The last data set considered in this paper is a brain cancer dataset
(\cite{Pomeroy:2002:1}). The total number of patients in this case is $n=42$, each represented
by $p=5597$ microarray gene expression features, covering $5$ different types of brain cancer. Using the packages R packages {\tt rrcov} and {\tt rrcovHD}, we first
computed the apparent error for each of the methods on all the datasets mentioned above.

\subsection{Comparisons of Methods on Real Data Using the Apparent Error}
Although the core of our work in this paper is focused on predictive performance as stated right
from our introduction, we devote this subsection to comparisons based
on the apparent (training) error, following from \cite{Todorov:2007:1} and
\cite{Pires:2010:1} who performed similar comparisons. At the very least, the computation (or attempt thereof)
of the apparent misclassification rates gives a rough idea of how well the method
might perform predictively, when they can be applied at all.

\begin{table}[!h]
\centering
\begin{tabular}{lrrrrrrr}
\toprule
$\frac{n}{p}$& \tiny{$\frac{145}{3}=48.33$} & \tiny{$\frac{79}{500}=0.158$} & \tiny{$\frac{180}{661}=0.272$} & \tiny{$\frac{197}{1000}=0.197$} & \tiny{$\frac{62}{2000}=0.031$} & \tiny{$\frac{72}{3571}=0.020$} & \tiny{$\frac{42}{5597}=0.0075$} \\
         &Diabetes&Prostate&Lymphoma&Lung&Colon&Leukemia&Brain \\ \toprule
 \bf{Classic} & 13.10 &   NA  &  NA   &  NA   &  NA    &  NA & NA    \\
 \bf{Linda}   & 10.35 &   NA  &  NA   &  NA   &  NA    &  NA & NA    \\
 \bf{PP}      & 27.58 & 29.11 & 55.00 & 21.83 &  11.29 & 5.55& 52.38  \\
 \bf{SIMCA}   & 11.72 & 35.44 &  9.44 &  5.58 &  9.68  &12.50& NA     \\ \bottomrule
\end{tabular}
\caption{Apparent misclassification rates
         for the classical and robust estimators under different scenarios}
\label{tab:1}
\end{table}

{\it As can be seen on Table \eqref{tab:1}, Linda (which is an R implementation of the MCD approach) and
classic LDA only work when $n/p$ is greater than 1, in this case on the diabetes data. PP and SIMCA
do handle HDLSS $(n/p < 1)$, with the exception that SIMCA fails when $n/p < 10^{-2}$. When they
 can both be applied, there is no clear winner between PP and SIMCA: on some datasets, PP outperforms
 SIMCA, but on other datasets, it is the other way round. This seems to indicate that besides the impact
of $n/p$, there is also the effect of the internal geometry of the data.}

\subsection{Comparisons of Methods on Real Data  Using the Average Test Error}
\begin{figure}[!h]
  \centering
  \epsfig{figure=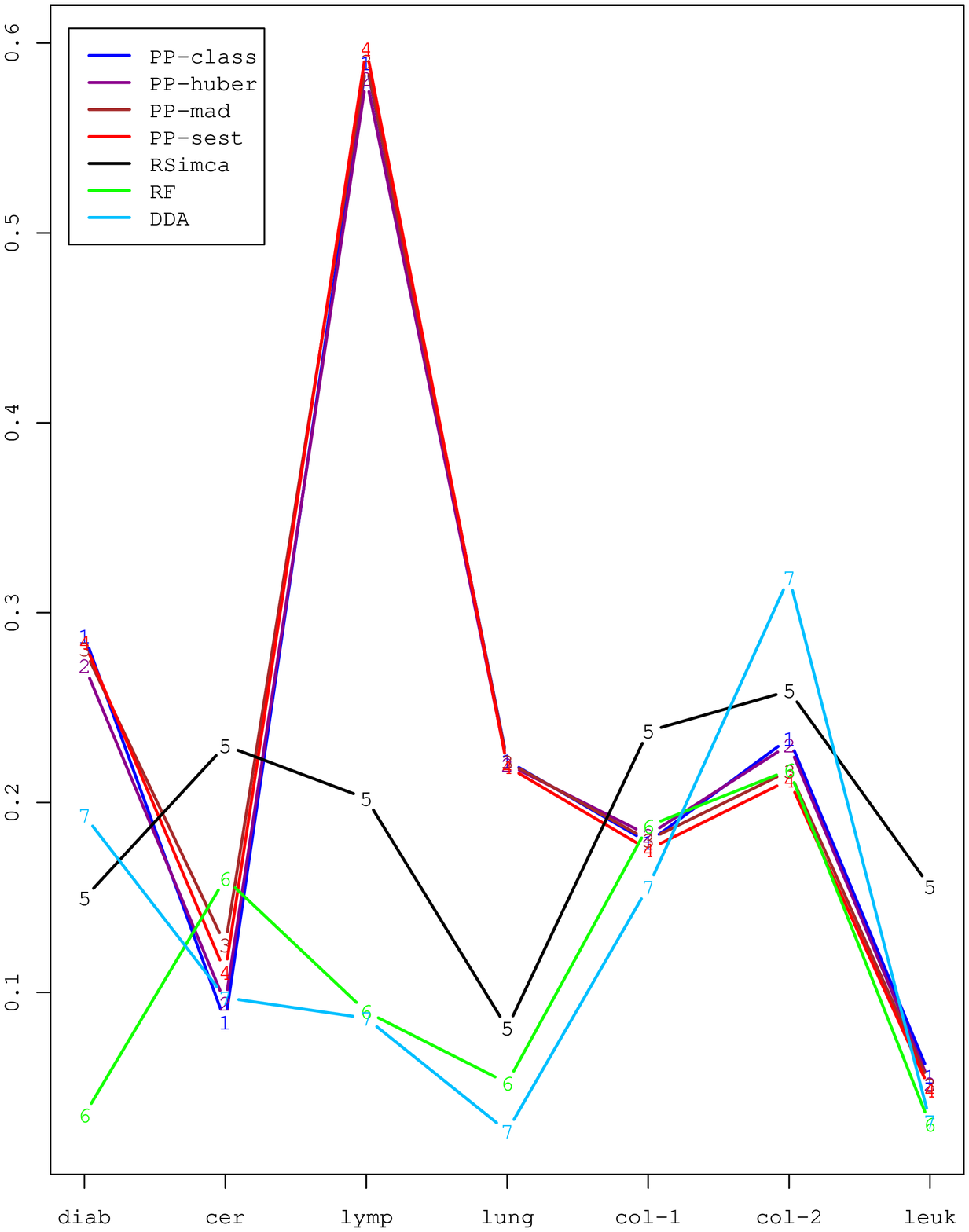, width=7cm,height=7cm}
  \epsfig{figure=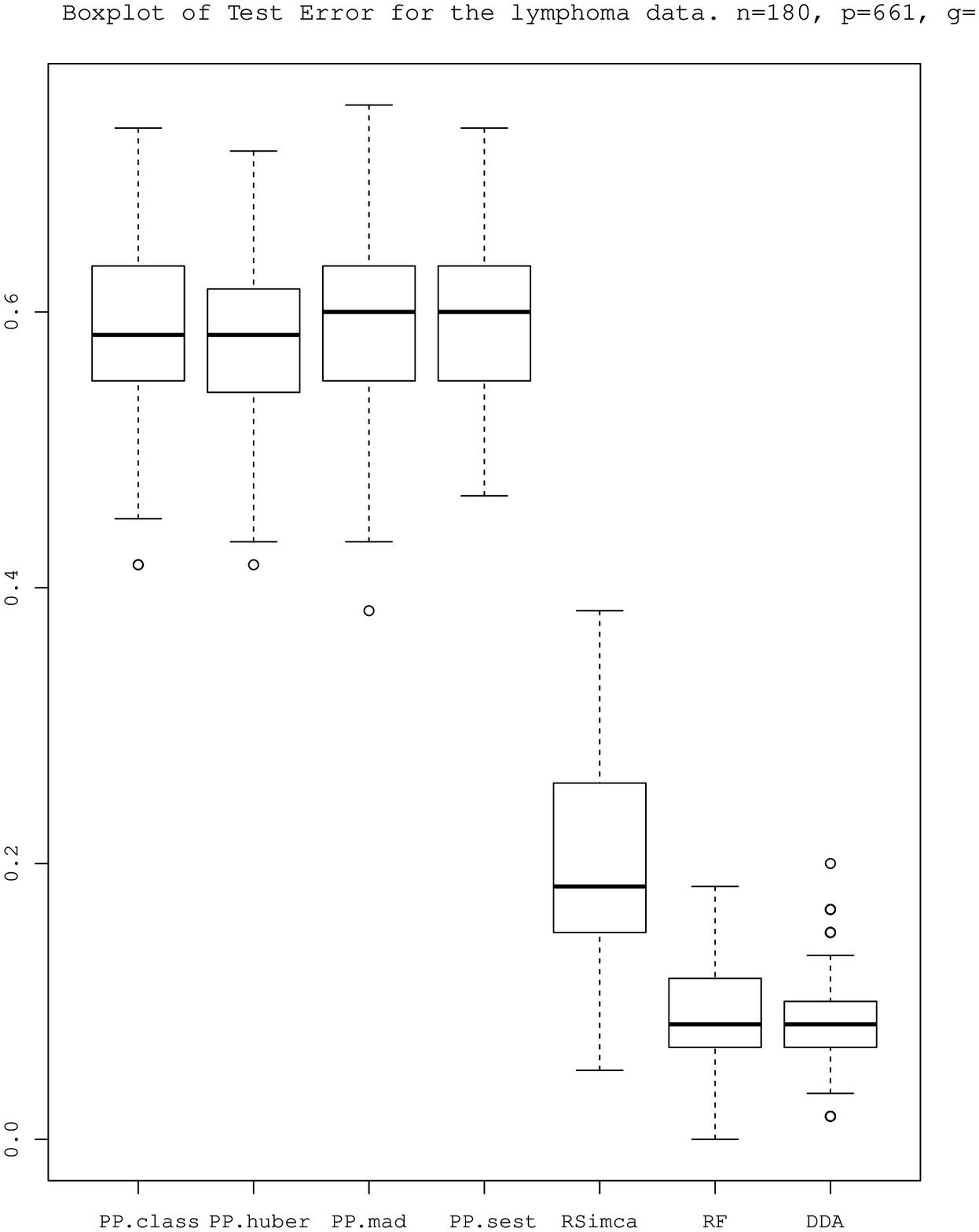, width=7cm,height=7cm}
  \caption{(left) Average test error as a function of $n/p$. The data sets appear on the x axis in decreasing order of $n/p$.
  The first one(diabetes) has $n/p=145/3$ and the last one (leukemia) has $n/p=72/3571$;
  (right) Average test error on the lymphoma data set for which $n/p=180/661$, and the number of classes is $g=3$.
  These box plots compare the predictive performances of all the 7 methods considered.}
  \label{fig:real:all:and:lymphoma:1}
\end{figure}

\noindent In the interest of comparing the predictive optimality of the techniques explored,
we now present the performances of 4 different variants of PP discrimination, against robust SIMCA,
RF and Diagonal Discrimination Analysis (DDA). Here, we use $R=200$ replications, and some of the results are
summarized in Table \eqref{tab:real:all:1} and Figure  \eqref{fig:real:all:and:lymphoma:1}. As revealed in Table \eqref{tab:real:all:1},
one of the most immediate remark
that emerges from our computations is the fact that SIMCA comes out as the worst in predictive performance 3 times,
which is far more than any other method. SIMCA also never comes out as best. We actually provide greater details about this mediocre aspect of SIMCA in the simulated data section later. Another noteworthy aspect of these computations is the fact that
DDA appears to perform very well, specifically emerging as the best 3 times (which is more times than any other technique).
Given the fact DDA operates under the strong assumptions of uncorrelatedness of the predictor variables, we are let
to infer that most of the datasets, especially those on which DDA has the best predictive performances, are inherently very high dimensional.
In fact, this line of thought is somewhat supported by the fact overall, PP, which relies on the existence of a lower dimensional projection of the data, is the most unstable of the all the methods involved as shown clearly by Figure \eqref{fig:real:all:and:lymphoma:1} where PP
depicts huge spikes corresponding to large prediction errors. As depicted on the right panel of Figure  \eqref{fig:real:all:and:lymphoma:1},
the lymphona data causes PP to fail miserably, probably because the intrinsic dimensionality of this data is so
high that the projections fail to find lower dimensional representations.
RF on the other hand provides what we perceive as the most stable of all the
performances: it can be seen in Table \eqref{tab:real:all:1} and Figure  \eqref{fig:real:all:and:lymphoma:1} that RF never comes last,  and
is usually best or second best, with no instability spikes. As a matter of fact, the right panel of Figure  \eqref{fig:real:all:and:lymphoma:1}
shows the superior performance of RF and DDA, and relatively good performance of SIMCA on the lymphoma data, whereas all the PP variants fail catastrophically.
\begin{table}[ht]
\centering
\begin{tabular}{lrrrrrrr}
  \hline
 & PP-class & PP-huber & PP-mad & PP-sest & RSIMCA & RF & DDA \\
  \hline
  {\bf diabetes} & ${\it 28.79}^{\ddag}$ & 27.21 & 28.07 & 28.44 & ${\sc 14.96}^{\dag}$ & ${\bf 3.54}^*$ & 19.32 \\
                 & (5.41) & (5.68) & (5.57) & (5.40) & (4.95) & (2.53) & (4.99) \\
  {\bf ceramic} & ${\bf  8.44}^{*}$ & ${\sc 9.44}^{\dag}$ & 12.50 & 11.06 & ${\it 23.00}^{\ddag}$ & 16.00 & 9.72 \\
                & (9.07) & (8.32) & (10.14) & (9.35) & (13.62) & (10.13) & (9.96) \\
  {\bf lymphoma} & 58.96 & 58.14 & 59.03 & ${\it 59.72}^{\ddag}$ & 20.22 & ${\sc 8.98}^{\dag}$ & ${\bf 8.64}^{*}$ \\
             & (6.02) & (5.95) & (6.36) & (5.63) & (7.29) & (3.98) & (3.55) \\
  {\bf lung} & ${\it 22.20}^{\ddag}$ & 22.02 & 22.15 & 21.90 & 8.12 & ${\sc 5.23}^{\dag}$ & ${\bf 2.69}^{*}$ \\
             & (4.18) & (4.18) & (4.28) & (4.09) & (3.37) & (2.53) & (1.69) \\
  {\bf colon-1} & 17.90 & 18.29 & ${\sc 18.02}^{\dag}$ & 17.52 & ${\it 23.76}^{\ddag}$ & 18.76 & ${\bf 15.55}^{*}$ \\
                & (7.73) & (7.99) & (7.68) & (8.07) & (10.34) & (8.96) & (6.78) \\
  {\bf colon-2} & 23.38 & 23.05 & ${\sc 21.67}^{\dag}$ & ${\bf 21.19}^{*}$ & 25.88 & 21.71 & ${\it 31.86}^{\ddag}$ \\
                & (8.46) & (8.47) & (9.05) & (9.71) & (11.82) & (9.19) & (15.27) \\
  {\bf leukemia} &  5.58 & 5.21 & 5.13 & 4.85 & ${\it 15.56}^{\ddag}$ & ${\bf 3.04}^{*}$ & ${\sc 3.21}^{\dag}$ \\
                 & (5.04) & (4.19) & (4.90) & (4.65) & (12.47) & (4.04) & (3.66) \\
   \hline
\end{tabular}
  \caption{Average test error along with the corresponding standard deviation in parentheses. The star (*) is used to indicate the method with the best predictive performance, while the double dagger ($\ddag$) indicates the worst predictive performance. The dagger ($\dag$) identifies the second best. The absence of prostate and brain is due to the fact that many methods explored could not even handle them. SIMCA for instance could not handle the brain cancer data set.}
  \label{tab:real:all:1}
\end{table}

\subsection{Comparison of Predictive Performances on Simulated Data}
Based on the computations performed earlier on real life HDLSS microarray gene expression datasets,
some patterns began to emerge, among which the overall stability and relatively strong predictive performance
  of RF, a method not structurally aimed at robustness. We also noticed that PP is a very unstable method,
  typically producing the worst performances on most data sets. Despite all these initial findings, we still do not have
  a general characterization of which aspects of the data drive the performances of each method.
  In this section, we use a thorough simulation study with various aspects of data characteristics,
  with the finality of determination what makes methods work well. Throughout this paper, all our simulated data will be generated from
the multivariate Gaussian whose density will be denoted by $\phi_p(\vx; \bfmu, \Sigma)$, and defined by
\begin{eqnarray}
   \label{eq:p:gauss}
    \phi_p(\vx; \bfmu, \Sigma) = \frac{1}{\sqrt{(2\pi)^p |\Sigma|}} \exp\left\{-\frac{1}{2}(\vx-\bfmu)^\top \Sigma^{-1}(\vx-\bfmu)\right\}.
\end{eqnarray}

Under the $\epsilon$-contamination regime, the class conditional density of $X$
in class $k$ is given by
\begin{eqnarray}
   \label{eq:2:1a}
   p(\vx | \bfmu, \Sigma, k,\epsilon, \eta, \kappa) =
  (1-\epsilon)\phi_p(\vx; \mu_k,\Sigma)+\epsilon \phi_p(\vx; \mu_k+\eta, \kappa\Sigma),
\end{eqnarray}
where $\eta$ represents the contamination of the location, while $\kappa$ captures the
level of contamination of the scatter matrix. Furthermore, in order to study the effect of the
correlation pattern, we simulate the data using a covariance matrix $\Sigma$ parameterized by
$\tau$ and $\rho$ and defined by
$$
\Sigma = \Sigma(\tau, \rho) = \tau\left(
                        \begin{array}{ccccc}
                          1 & \rho & \cdots & \cdots &\rho \\
                          \rho & 1 & \rho & \cdots & \rho \\
                          \vdots & \ddots & \ddots & \ddots &  \vdots\\
                          \rho & \ddots & \rho & 1 &  \rho\\
                          \rho & \cdots & \cdots & \rho & 1 \\
                        \end{array}
                      \right) = \tau[(1-\rho)\bfI_p + \rho \bf1_p  \bf1_p^\top]
$$
where $\bfI_p$ is the $p$-dimensional identity matrix, while $\bf1_p$ is
$p$-dimensional vector of ones. In fact, a more general covariance matrix to help
better gauge the effect of correlation on the robust classification methods would be
$\tau\Sigma$, where $\Sigma= (\sigma_{ij})$, with $\sigma_{ij}=\rho^{|i-j|}$, that is,
$$\Sigma = \Sigma(\tau, \rho) = \tau\left(
                        \begin{array}{ccccc}
                          1 & \rho & \cdots & \rho^{p-2} &\rho^{p-1} \\
                          \rho & 1 & \rho & \cdots & \rho^{p-2} \\
                          \vdots & \ddots & \ddots & \ddots &  \vdots\\
                          \rho^{p-2} & \ddots & \rho & 1 &  \rho\\
                          \rho^{p-1} & \rho^{p-2} & \cdots & \rho & 1 \\
                        \end{array}
                      \right).
$$
For simplicity however, we use the first $\Sigma$ with $\tau=1$ throughout this paper. For the remaining
parameters, we use
$\epsilon \in \{0, 0.05, 0.15\}$, $\kappa \in \{9, 25, 100\}$,
$G \in \{2, 3\}$ and $\rho \in \{0, 0.25, 0.75\}$ and $p \in \{10, 100, 1000\}$. As the vector
of $\epsilon$ values shows, we consider 3 different levels of contamination, namely
no contamination, mild contamination and strong contamination.

\subsubsection{Uncontaminated Data}
We first consider the performances of the techniques under an uncontaminated regime, i.e. $\epsilon=0$.
Our first simulation on under this regime looks at combination where the number of classes is $G=2$
and than investigate the effect of $\rho$ and $p$ (input space dimension). As the plots all reveal,
PP appears to perform very well (usually outperforming all the other methods) whenever
intrinsic dimensionality of the data is low (captured by $\rho$ very high) and the number of classes is $2$ (Binary classification task).

\begin{table}[!ht]
\centering
\begin{tabular}{rrrrrrrr}
  \hline
 & PP-class & PP-huber & PP-mad & PP-sest & RSimca & RF & DDA \\
  \hline
  10 & 58.26 & 59.22 & 61.19 & 61.26 & 41.11 & 45.11 & 43.26 \\
  100 & 54.11 & 50.89 & 53.33 & 55.63 & 40.44 & 42.00 & 43.44 \\
  1000 & 44.44 & 35.26 & 35.74 & 35.96 & 30.81 & 28.70 & 35.89 \\
  2000 & 44.17 & 34.00 & 34.28 & 34.67 & 30.89 & 32.44 & 38.50 \\
   \hline
\end{tabular}
  \caption{Average test error on the uncontaminated simulated  data with $g=3$ and
  $\rho=0.75$. We herein reveal for each of the 7 methods, the effect of the input space dimension $p$ on the average test error.}
  \label{tab:sim:zero:23}
\end{table}

\begin{figure}
  \centering
  \includegraphics[width=5.5cm,height=5.5cm]{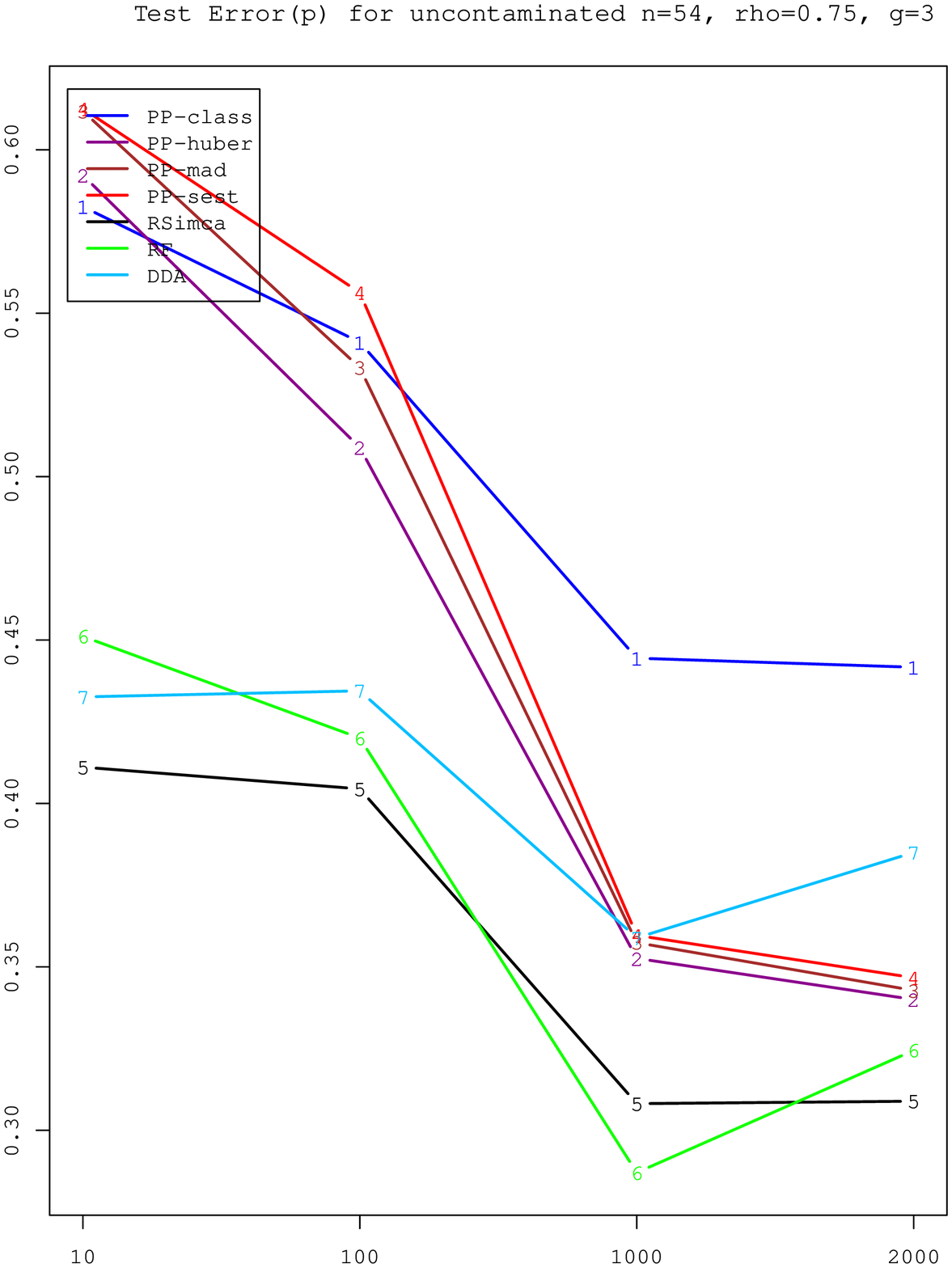}
  \includegraphics[width=5.5cm,height=5.5cm]{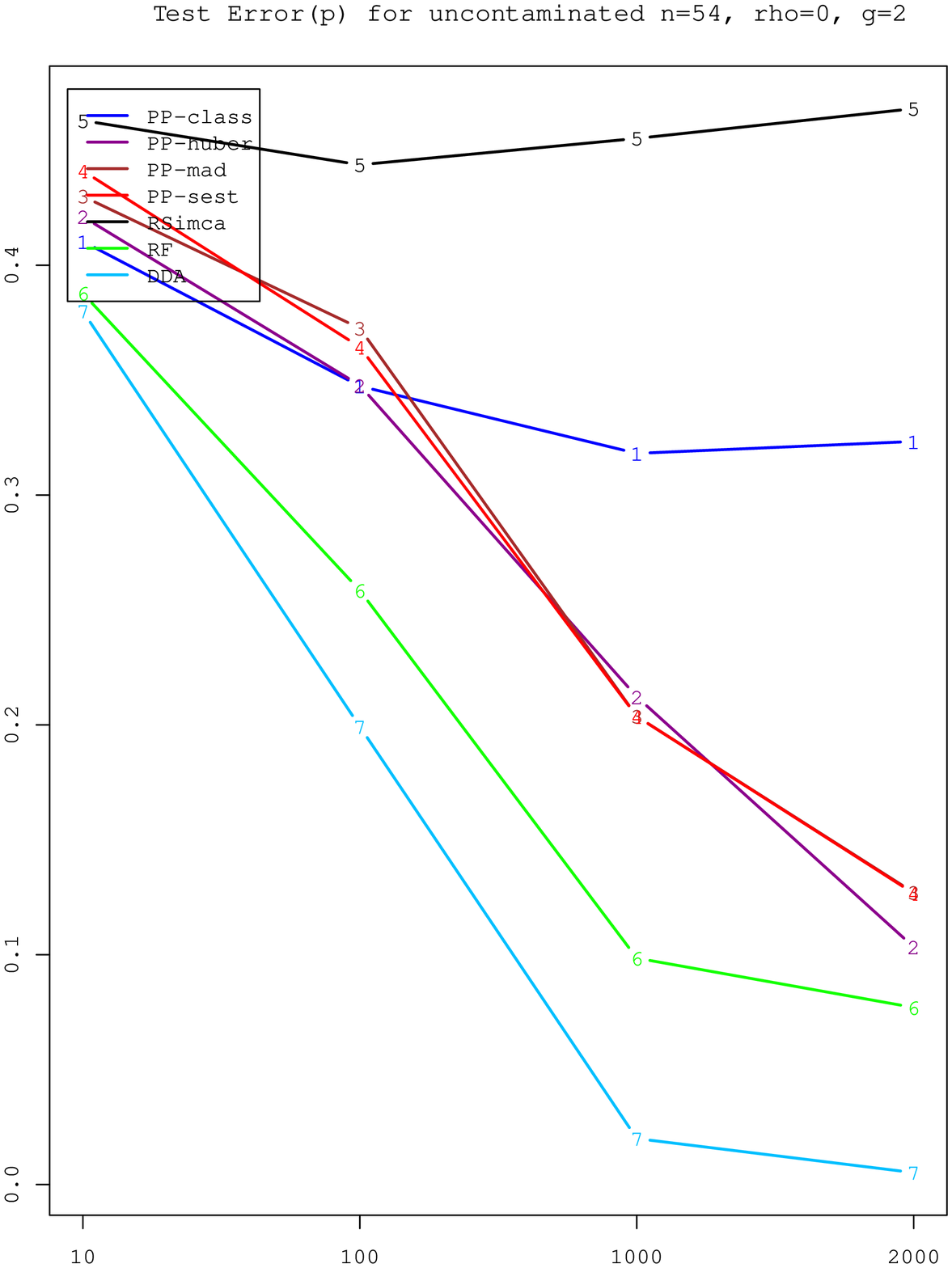}
  \includegraphics[width=5.5cm,height=5.5cm]{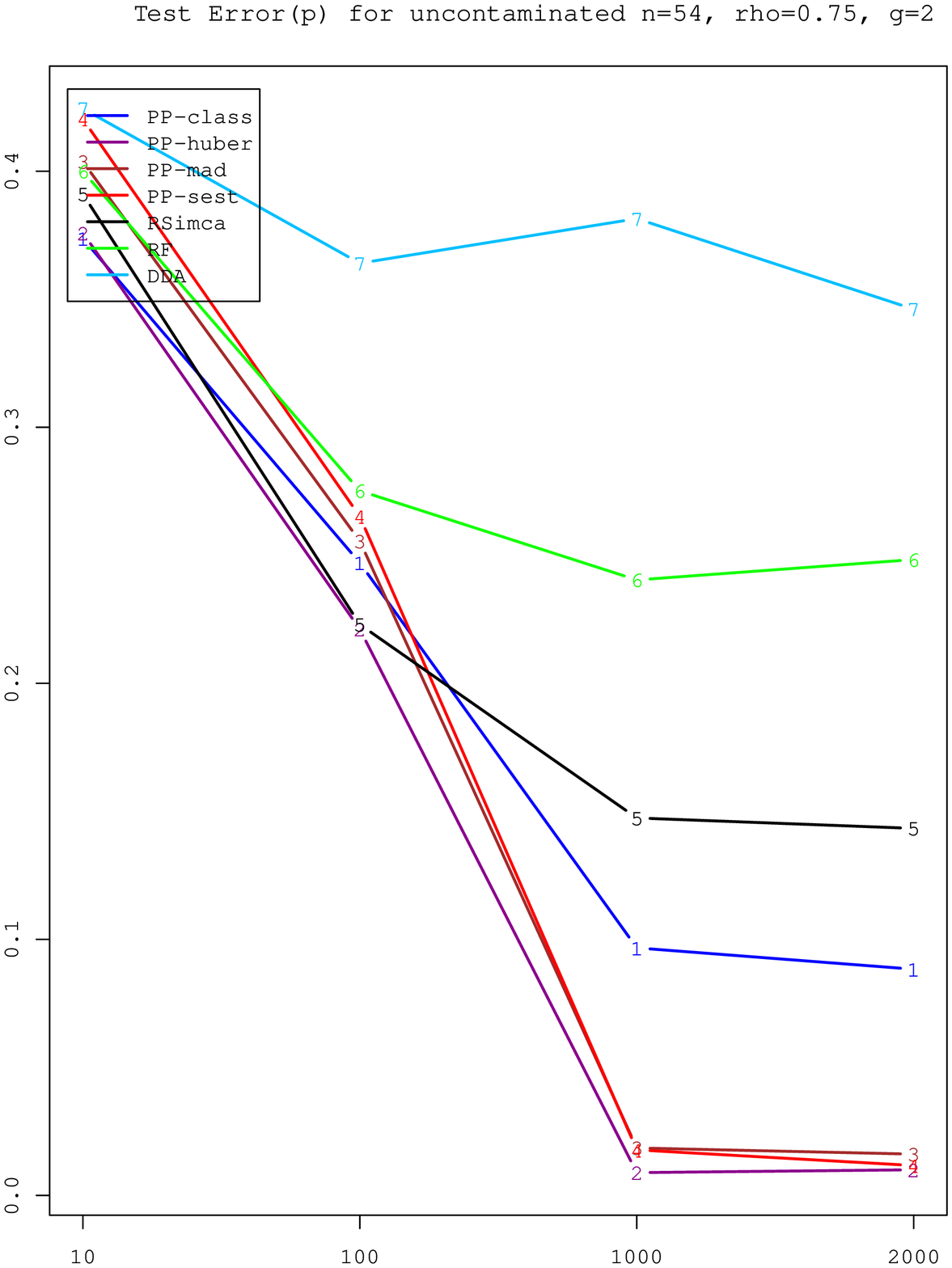}
  \caption{(left) Average test error on the uncontaminated simulated  data with $g=3$ and
  $\rho=0.75$. We herein reveal for each of the 7 methods, the effect of the input space dimension $p$ on the average test error.
  (center) Average test error on the uncontaminated simulated  data with $g=2$ and
  $\rho=0$. We herein reveal for each of the 7 methods, the effect of the input space dimension $p$ on the average test error.
  (right) Average test error on the uncontaminated simulated  data with $g=2$ and
  $\rho=0.75$. We herein reveal for each of the 7 methods, the effect of the input space dimension $p$ on the average test error.}
  \label{fig:sim:zero:all}
\end{figure}

Table  \eqref{tab:sim:zero:23} clearly reveals that PP does not work well ın multi-categorical classification.
With the number of classes just equal to $3$, PP yields the worst predictive performance, regardless of
the value of the overall correlation among the variables. As we discuss much later, PP
typically performs well in binary classification when $\rho$ is relatively large. But clearly, as depicted
in Table  \eqref{tab:sim:zero:23}, PP, at least in the implementation used here does NOT handle multiclass tasks well,
even when the data is potentially representable in a lower dimensional space (large $\rho$). It is important to
emphasize that this behavior of PP noticed here on uncontaminated data carries over to contaminated at various
rates of contamination. Figure \eqref{fig:sim:zero:all} clearly shows different scenarios of predictive performances under
different data characteristics when the data has no contamination. The left panel corresponds to the case when the number of classes is 3
and the correlation among variables is very high. In this case, RF and SIMCA emerge as the best whereas PP fails. In the center,
we see another excellent performance of RF. wıth DDA emerging as the best while PP fails, this time due to the fact that $\rho=0$.
The right panel highlight the ideal conditions for PP, namely binary classification along with large correlation.

\subsubsection{Effect of Mild Contamination}
\begin{figure}[!htbp]
  \centering
  \includegraphics[width=5.5cm,height=5.5cm]{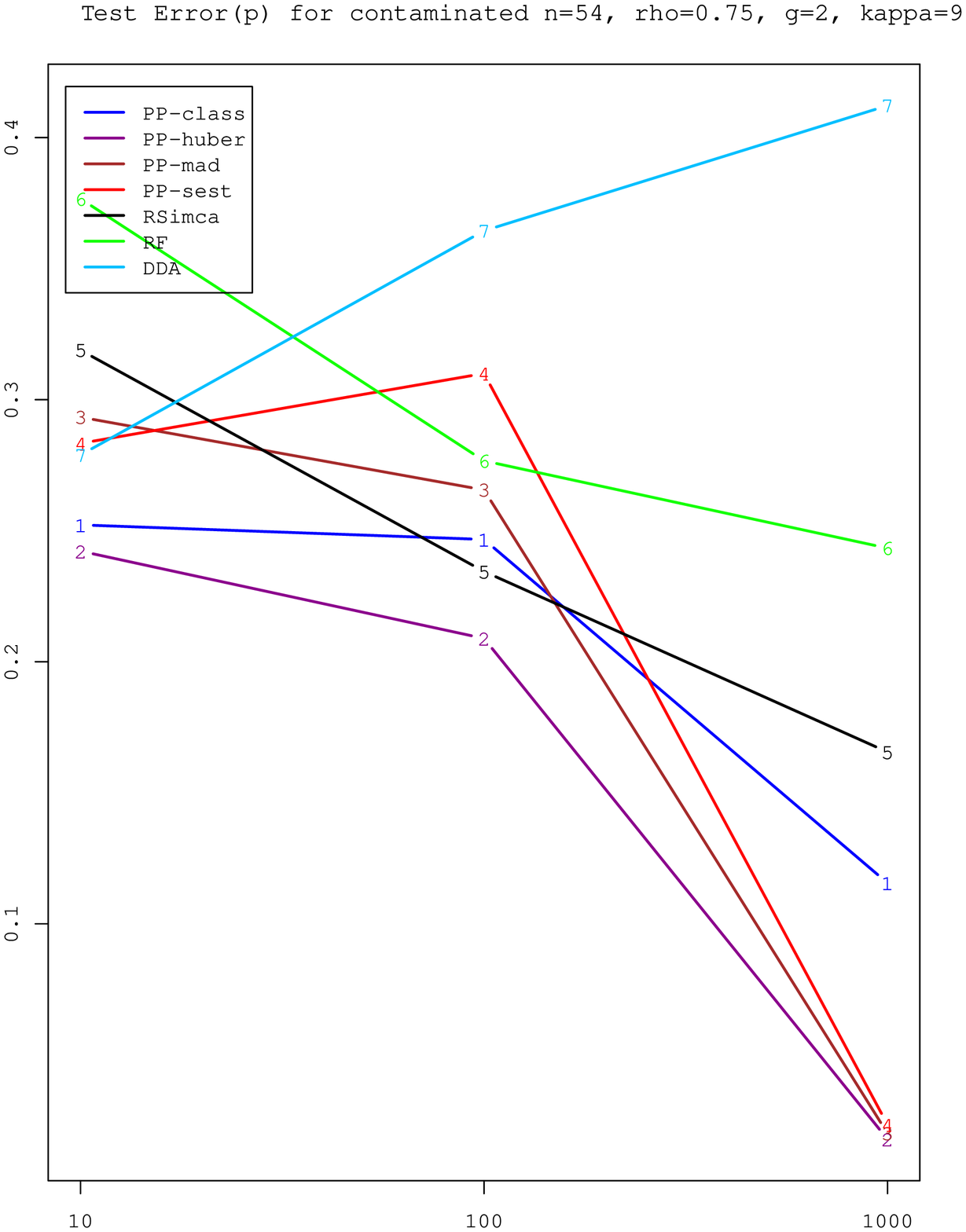}
  \includegraphics[width=5.5cm,height=5.5cm]{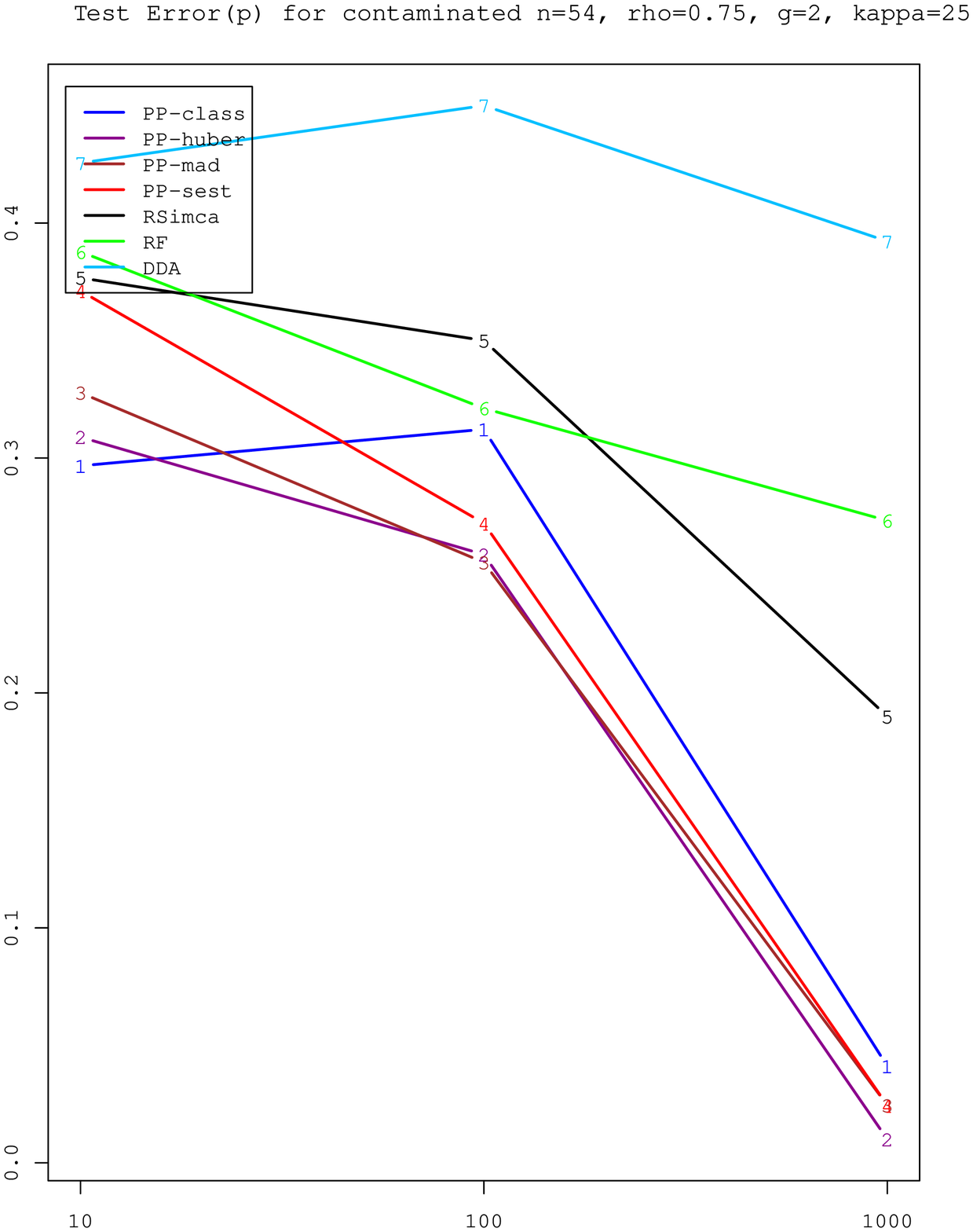}
  \includegraphics[width=5.5cm,height=5.5cm]{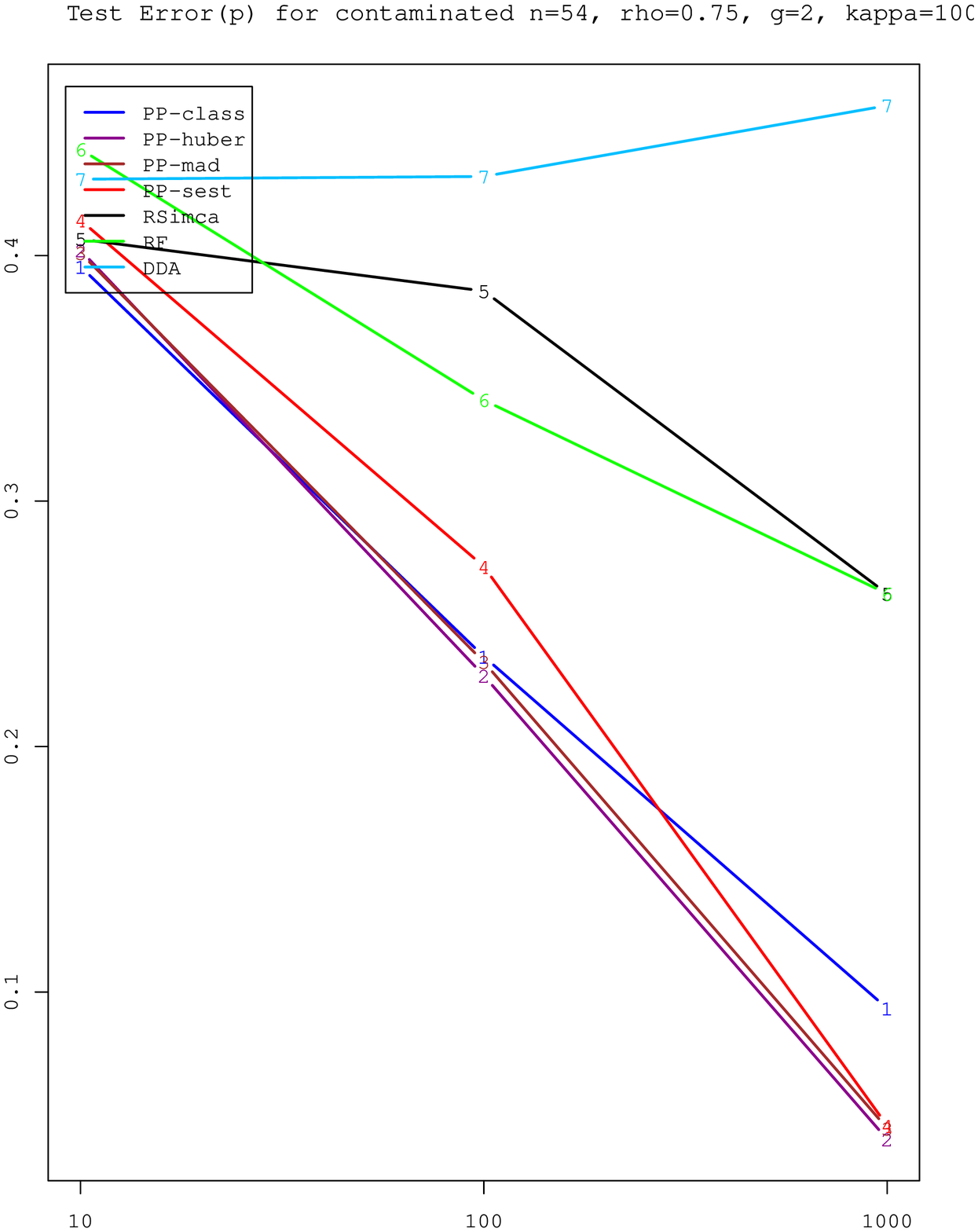}
  \caption{Average test error on the mild contaminated simulated  data with $g=2$ and
  $\rho=0.75$. We herein reveal for each of the 7 methods, the effect of the input space dimension $p$ on the average test error.}
  \label{fig:sim:mild:pp:1}
\end{figure}

We now consider the performances of the techniques under a mildly contaminated regime, i.e. $\epsilon=0.05$.
Our first simulation on under this regime looks at combination where the number of classes is $g=2$
and than investigate the effect of $\rho$ and $p$ (input space dimension) and $\kappa$.
Figure  \eqref{fig:sim:mild:pp:1} reveals what we anticipated earlier, namely that PP performs well,
typically emerging as the best predictive technique, when g=2 (binary classification) and $\rho$ is large (intrinsically
lower dimensionality of the data). In fact, under these two PP-favorable conditions, PP yields the best performance
regardless of the size $\kappa$ of contamination.

\begin{figure}[!htbp]
  \centering
  \includegraphics[width=5.5cm,height=5.5cm]{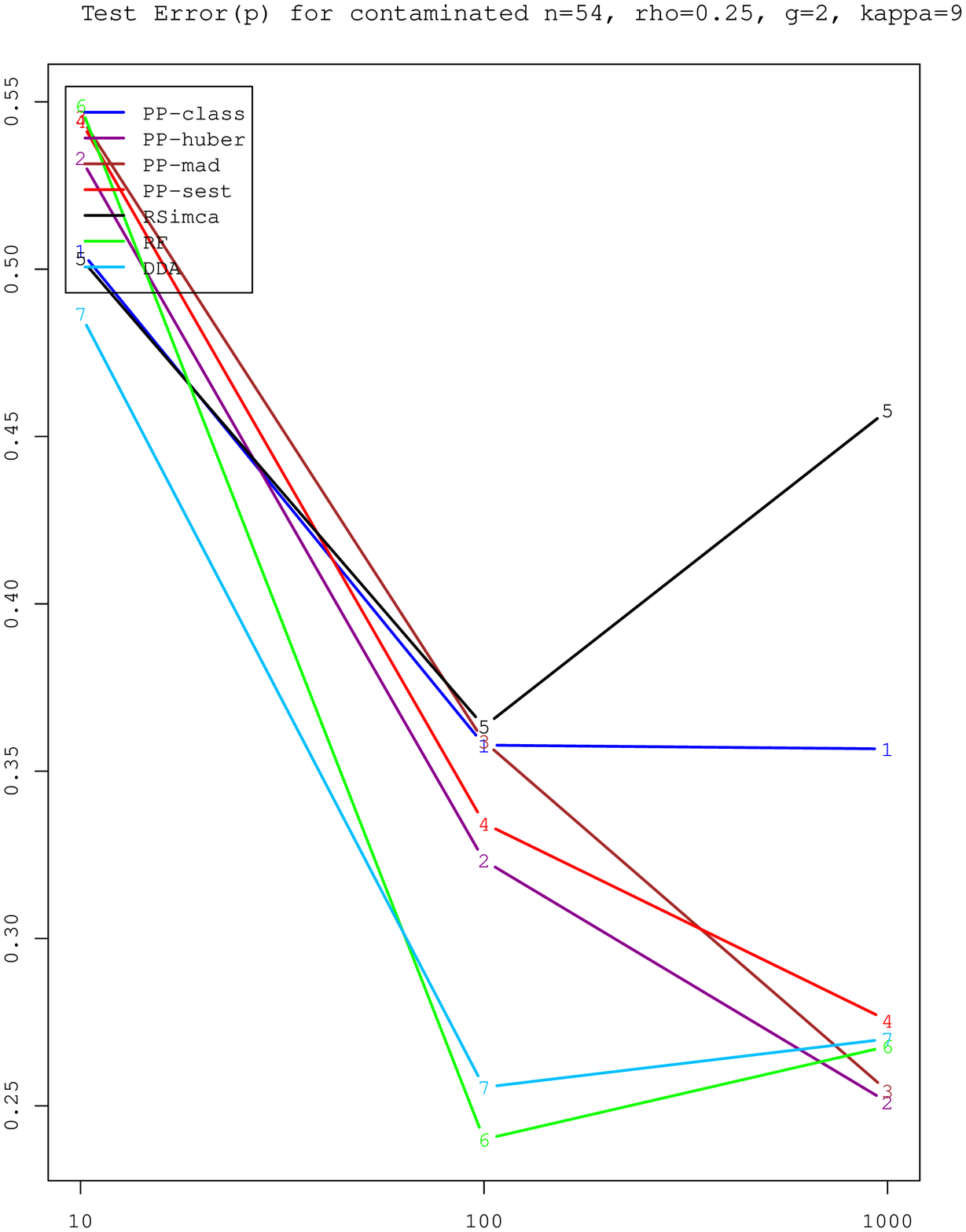}
  \includegraphics[width=5.5cm,height=5.5cm]{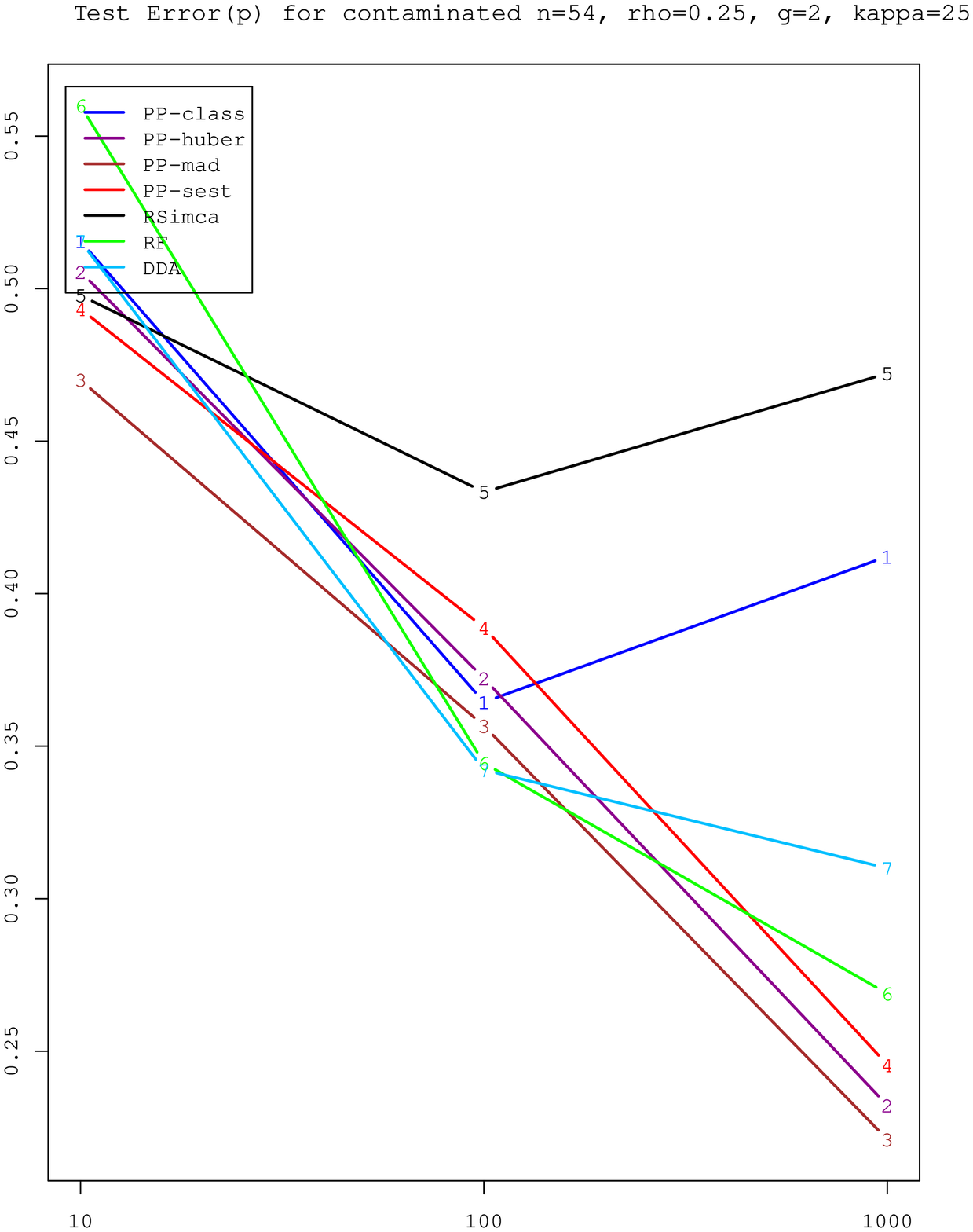}
  \includegraphics[width=5.5cm,height=5.5cm]{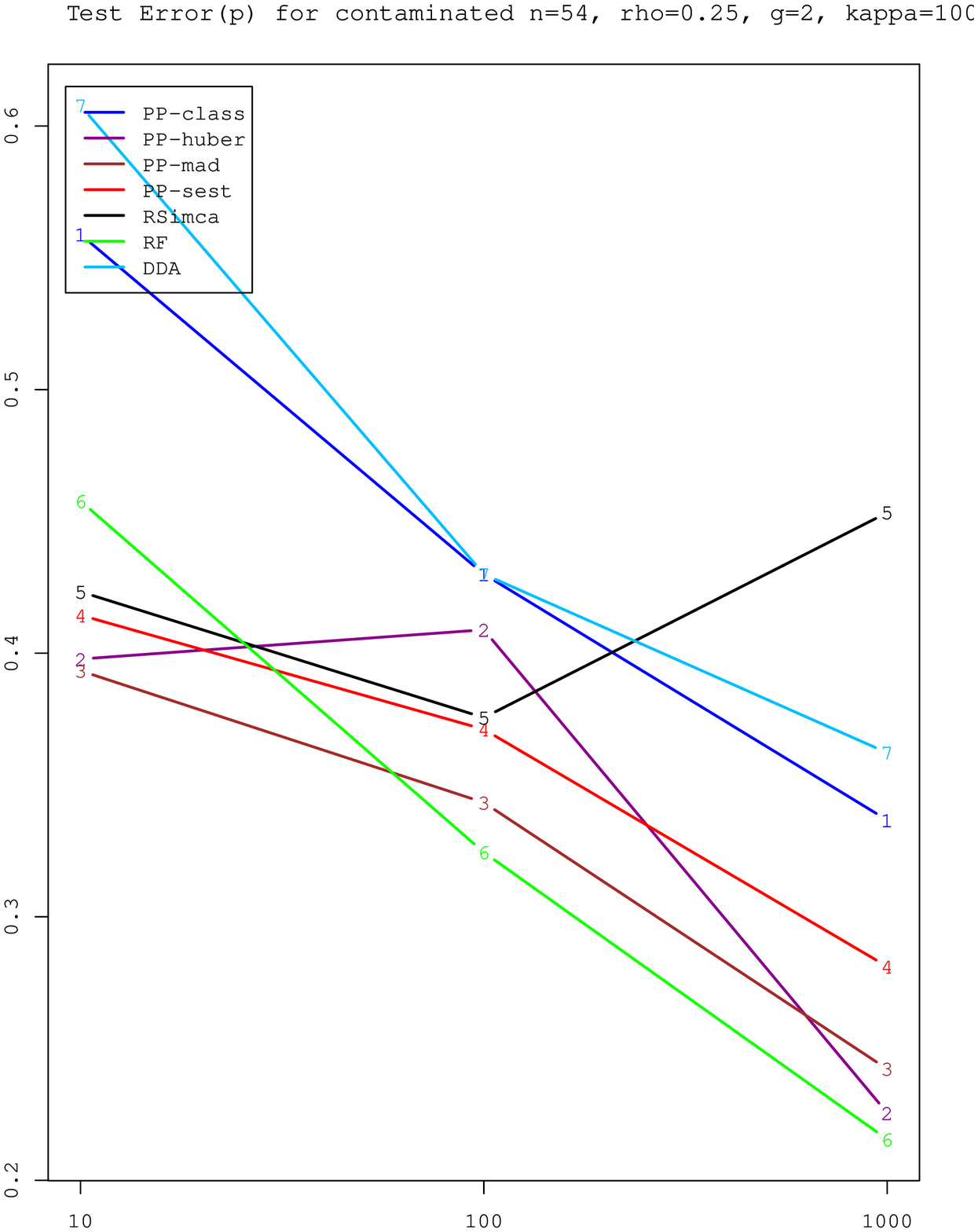}
  \caption{Average test error on the mild contaminated simulated  data with $g=2$ and
  $\rho=0.25$. We herein reveal for each of the 7 methods, the effect of the input space dimension $p$ on the average test error.}
  \label{fig:sim:mild:pp:2}
\end{figure}

\begin{figure}[!htbp]
  \centering
  \includegraphics[width=5.5cm,height=5.5cm]{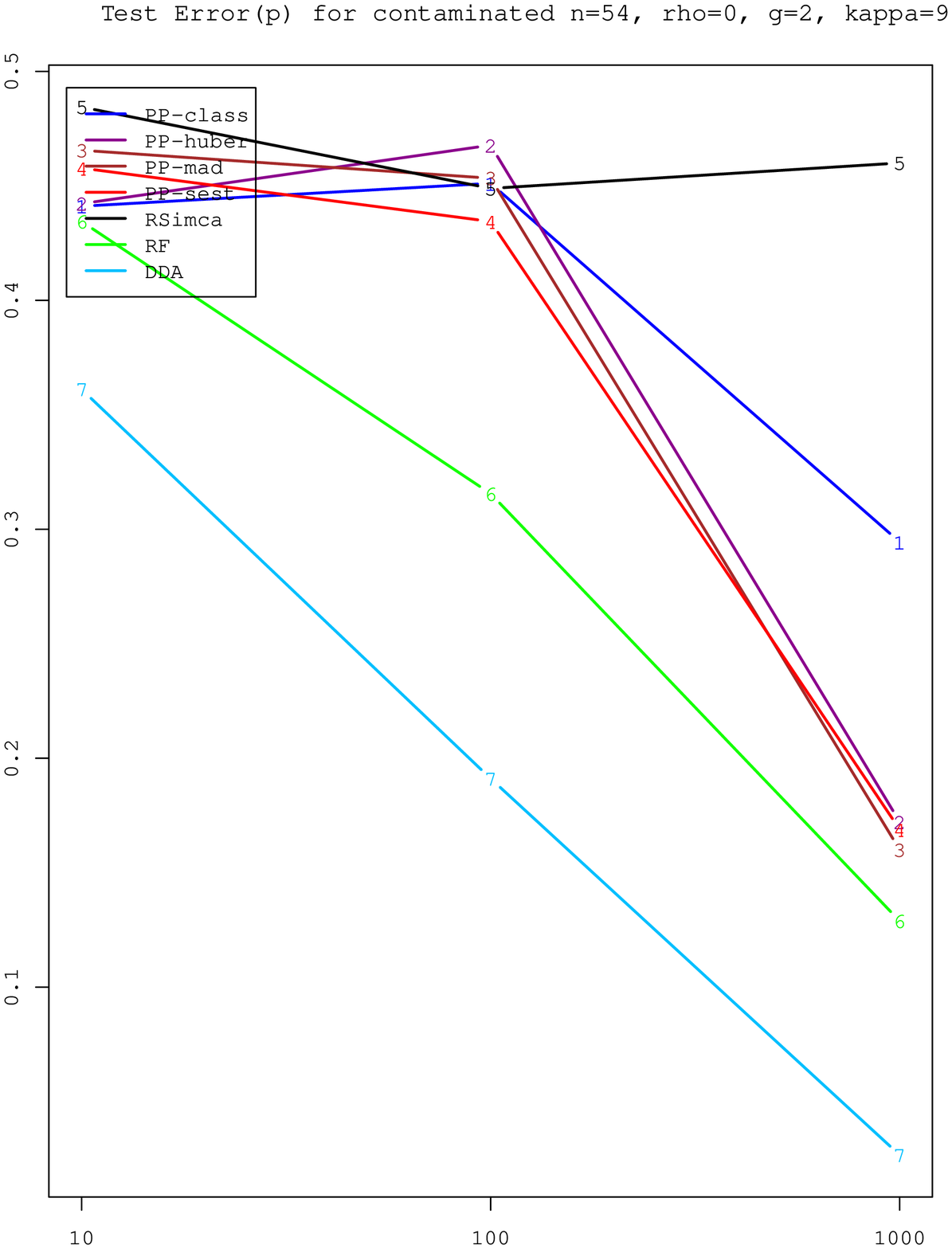}
  \includegraphics[width=5.5cm,height=5.5cm]{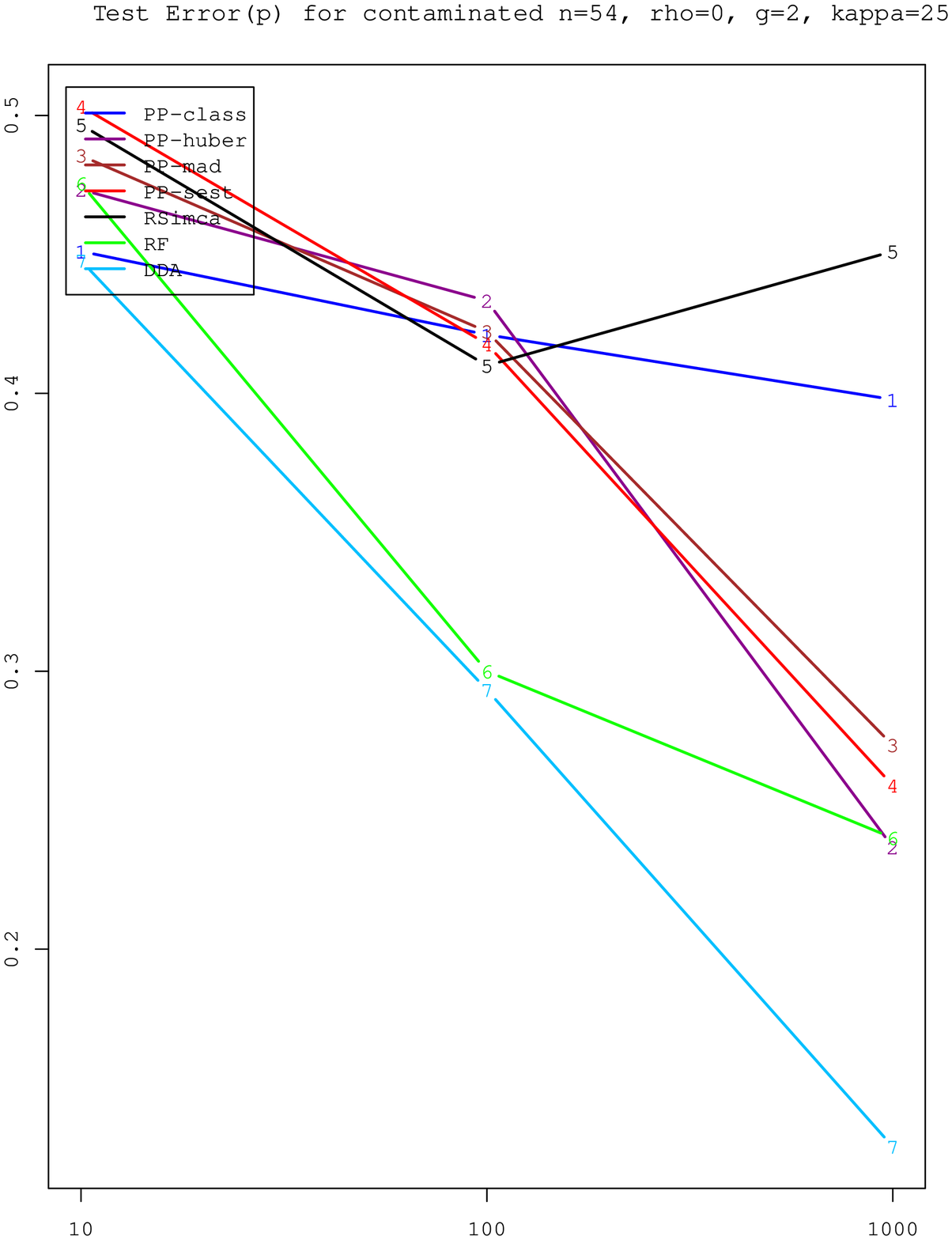}
  \includegraphics[width=5.5cm,height=5.5cm]{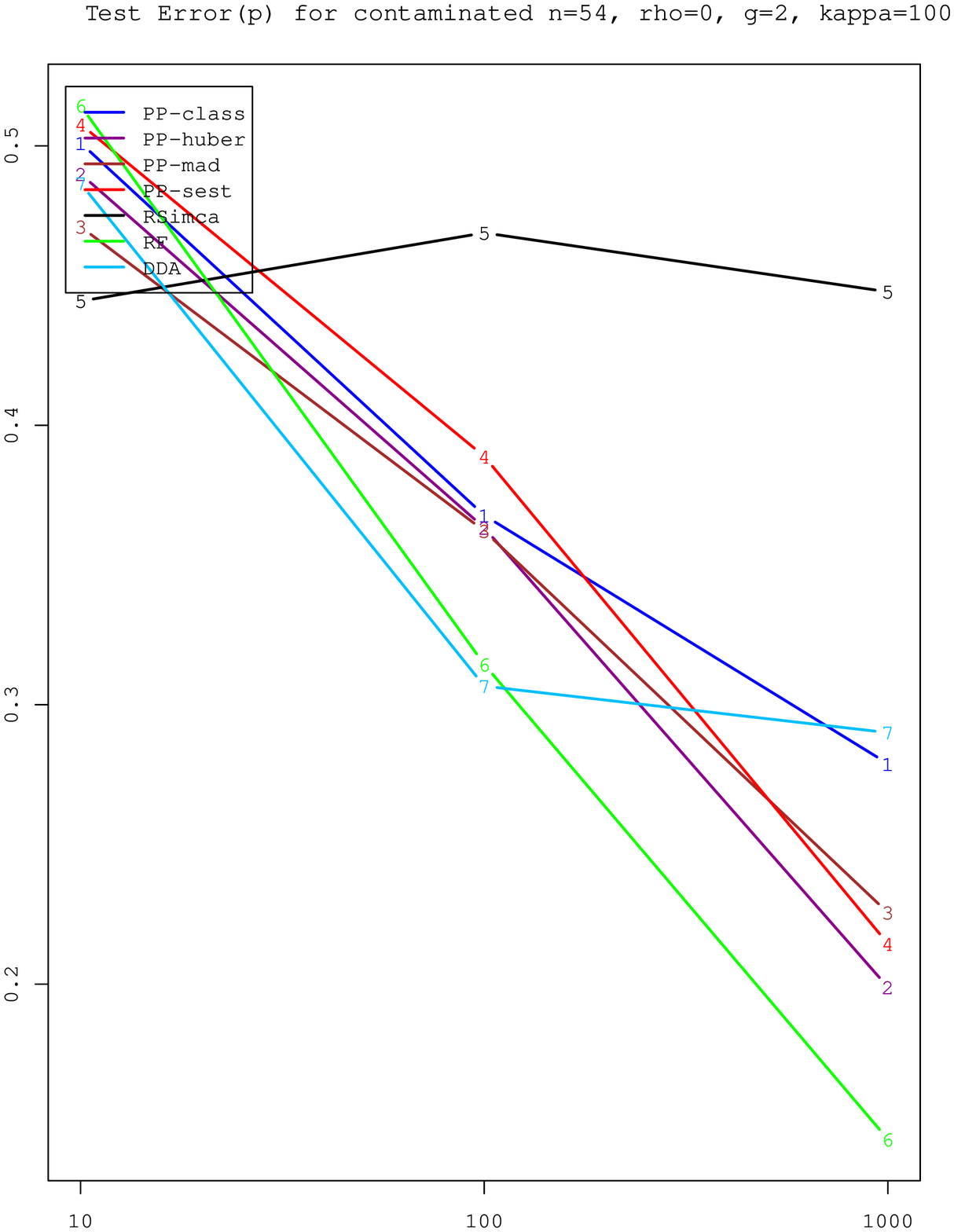}
  \caption{Average test error on the mild contaminated simulated  data with $g=2$ and
  $\rho=0.0$. We herein reveal for each of the 7 methods, the effect of the input space dimension $p$ on the average test error.}
  \label{fig:sim:mild:pp:3}
\end{figure}

\begin{figure}[!htbp]
  \centering
  \includegraphics[width=5.5cm,height=5.5cm]{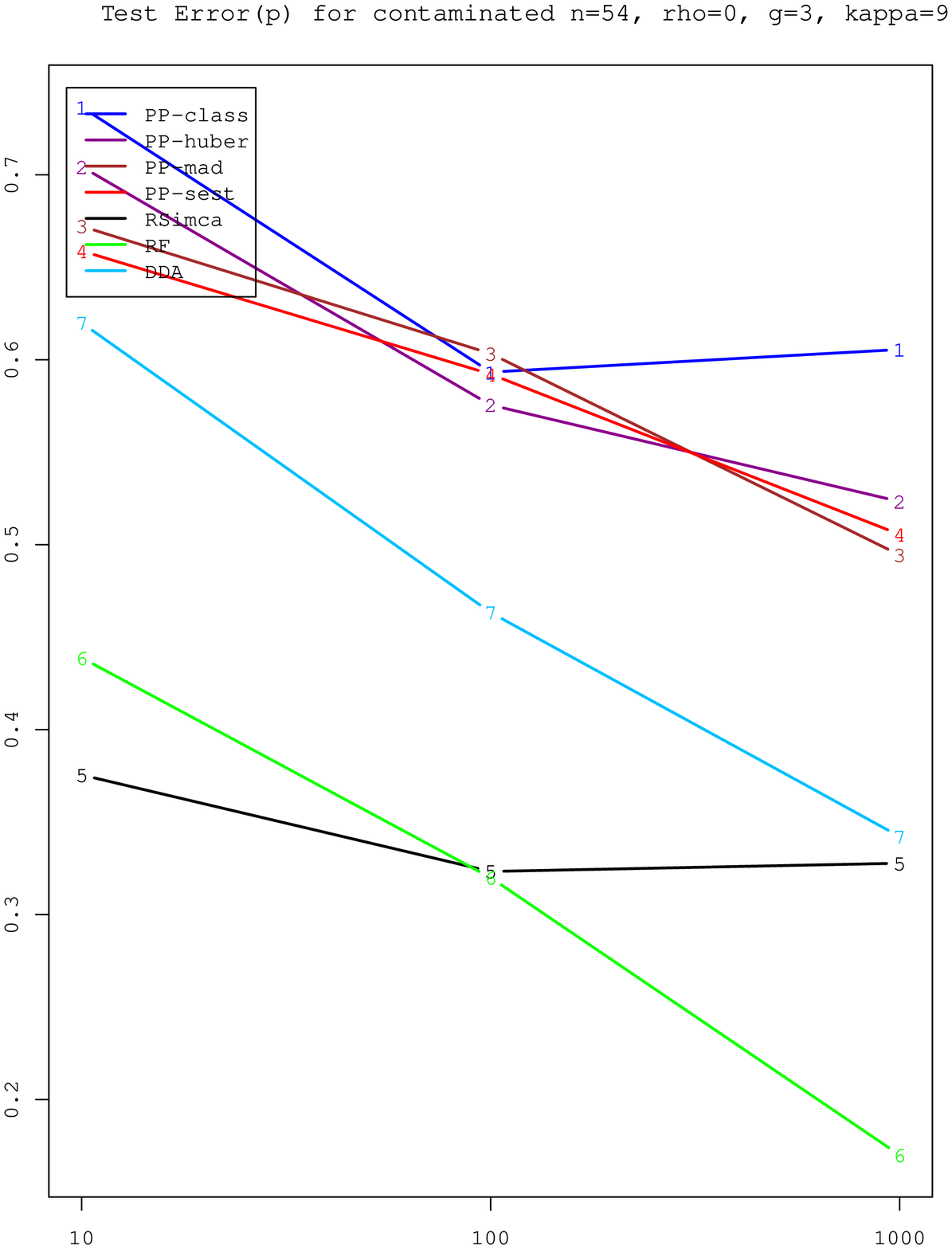}
  \includegraphics[width=5.5cm,height=5.5cm]{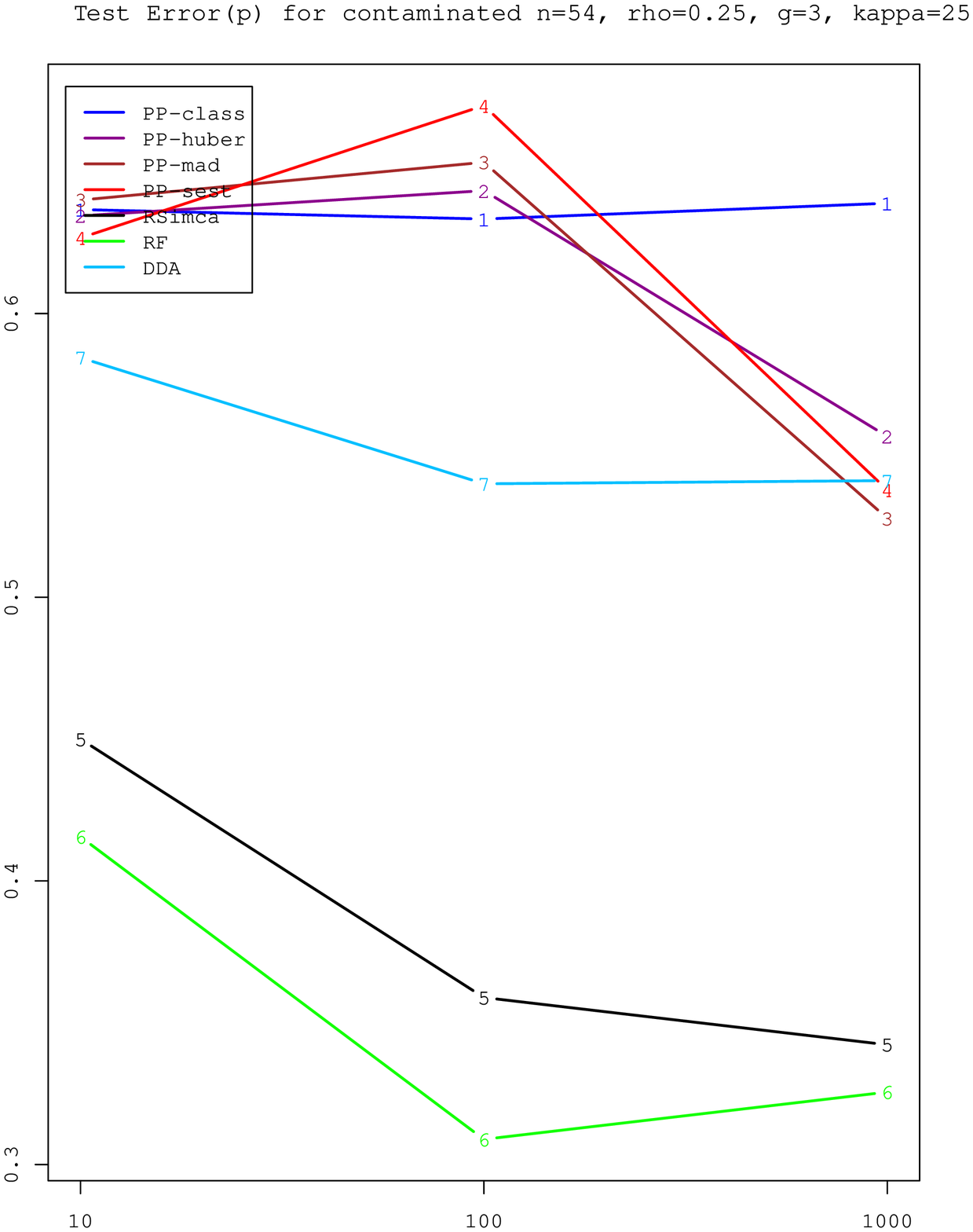}
  \includegraphics[width=5.5cm,height=5.5cm]{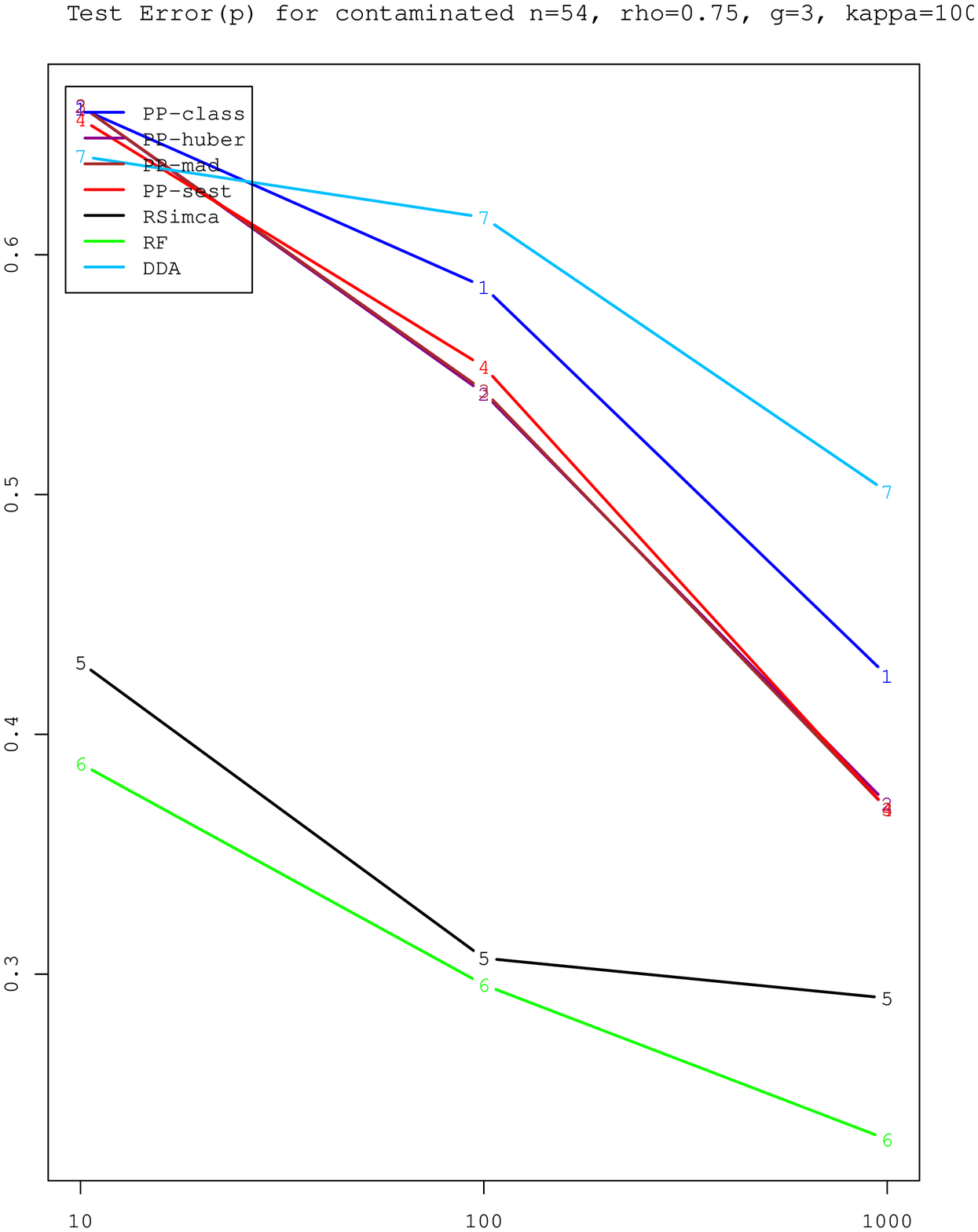}
  \caption{Average test error on the mild contaminated simulated  data with $g=3$.  We herein reveal for each of the 7 methods, the effect of the input space dimension $p$ on the average test error.}
  \label{fig:sim:mild:pp:4}
\end{figure}

Also noteworthy here is the fact that DDA fails for all the values
of $\kappa$ when $\rho$ is large. Finally, we also notice that RF does reasonably well,
while SIMCA gets progressively worse as $\kappa$ increases.

\subsubsection{Effect of Strong Contamination}
We now consider the performances of the techniques under a strongly contaminated regime, i.e. $\epsilon=0.15$.
Our first simulation under this regime looks at combination where the number of classes is $g=2$
and then investigates the effect of $\rho$ and $p$ (input space dimension) and $\kappa$.

\begin{figure}[!htbp]
  \centering
  \includegraphics[width=5.5cm,height=5.5cm]{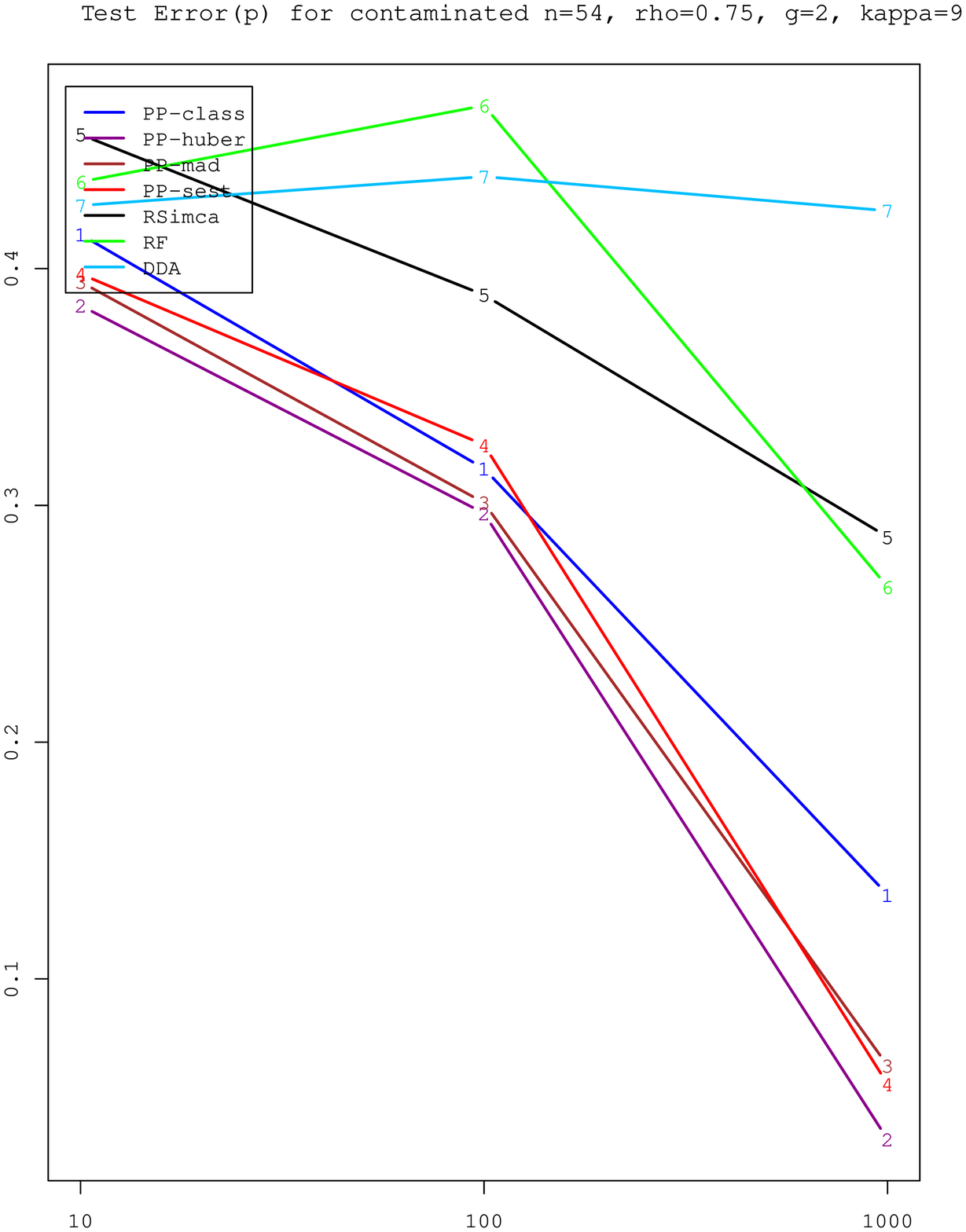}
  \includegraphics[width=5.5cm,height=5.5cm]{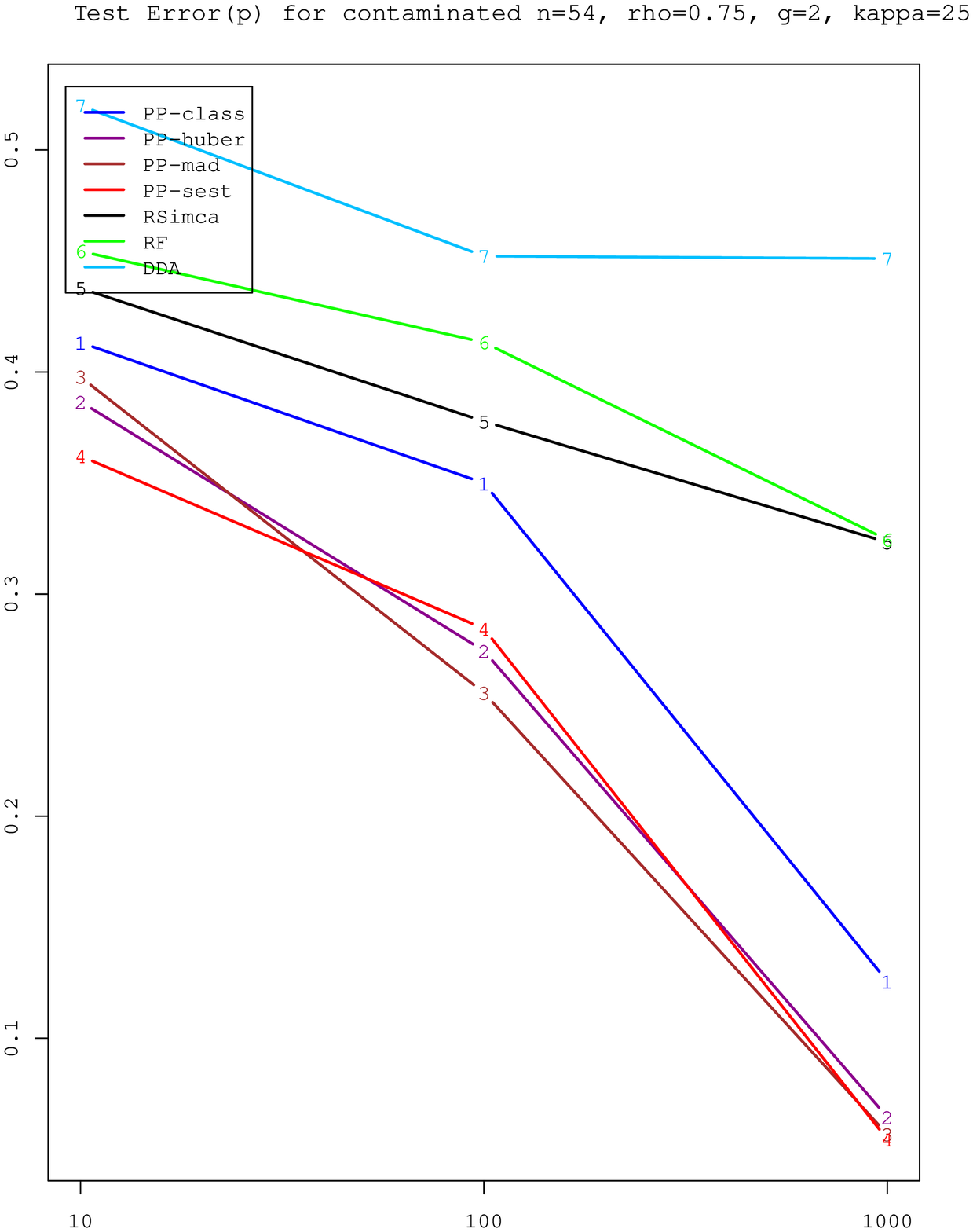}
  \includegraphics[width=5.5cm,height=5.5cm]{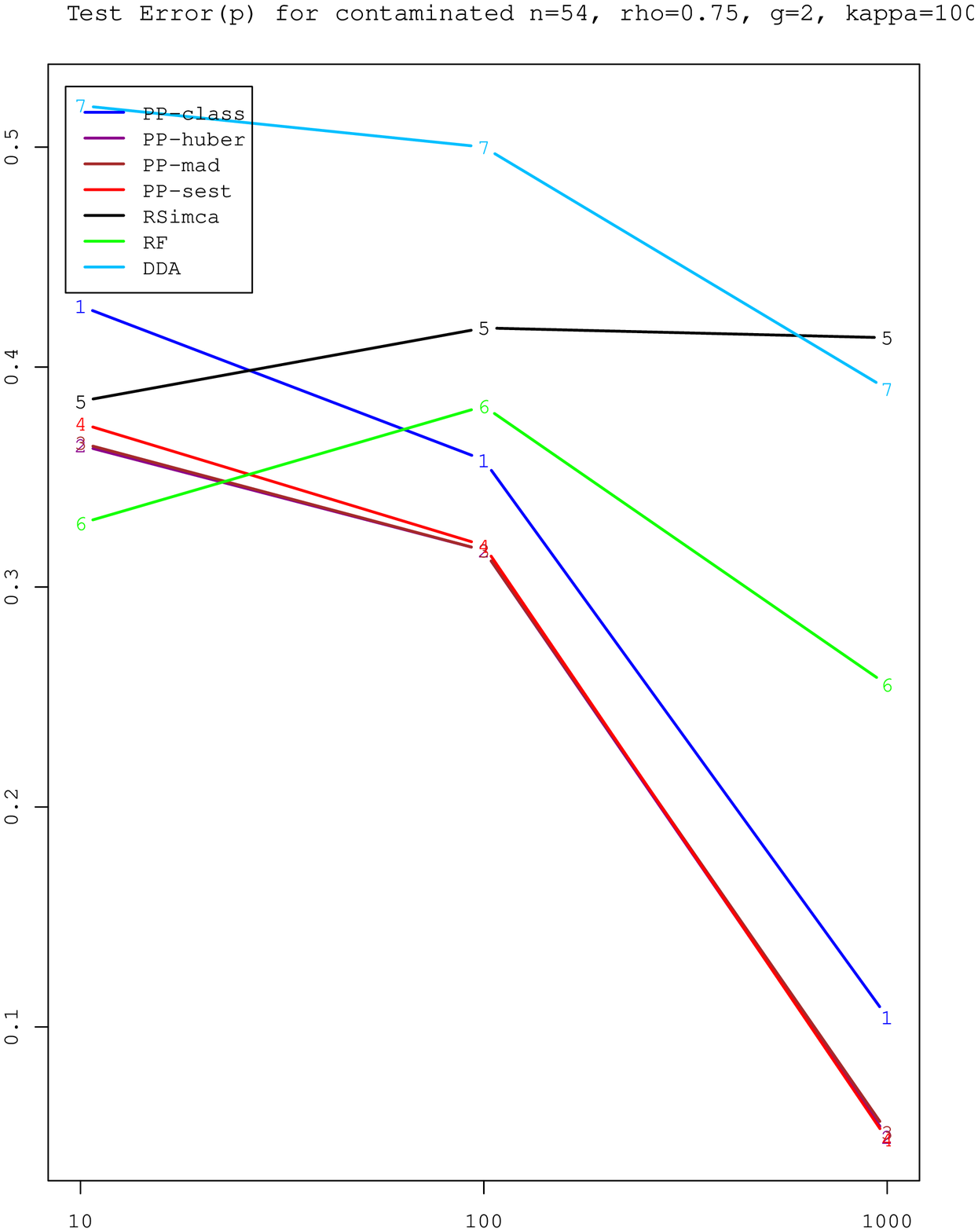}
  \caption{Average test error on the strongly contaminated simulated  data with $g=2$ and
  $\rho=0.75$. We herein reveal for each of the 7 methods, the effect of the input space dimension $p$ on the average test error.}
  \label{fig:sim:strong:pp:1}
\end{figure}

\begin{figure}[!htbp]
  \centering
  \includegraphics[width=5.5cm,height=5.5cm]{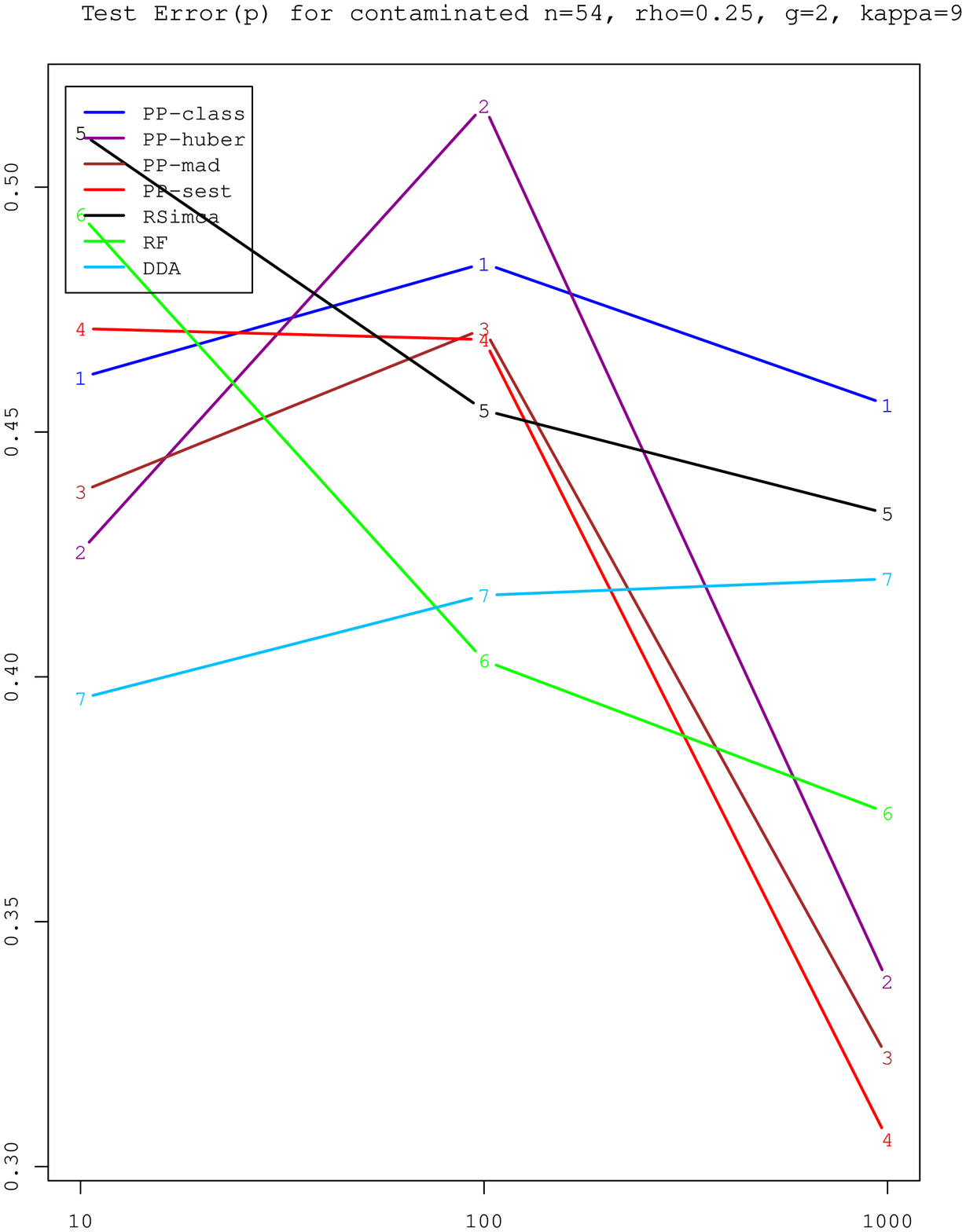}
  \includegraphics[width=5.5cm,height=5.5cm]{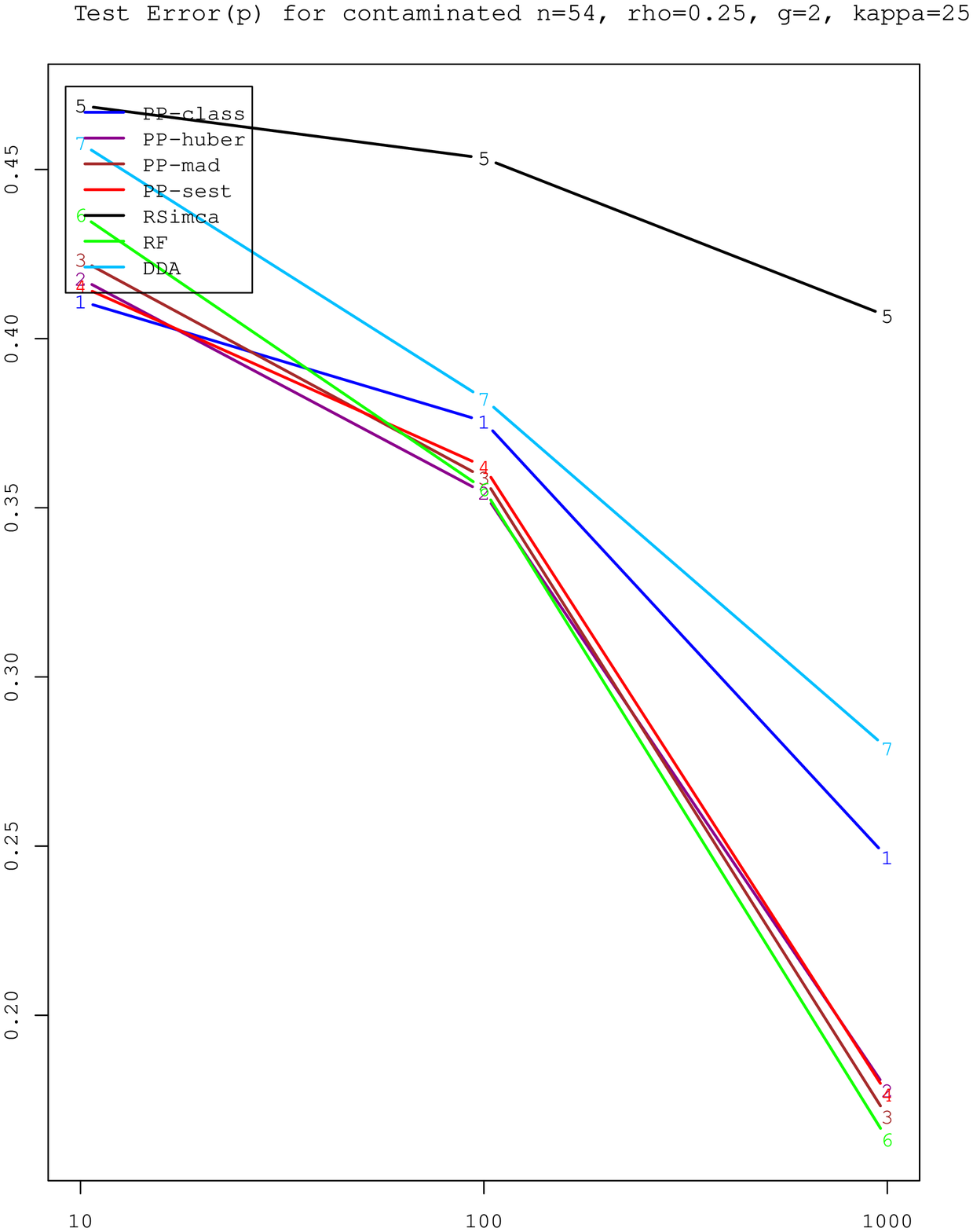}
  \includegraphics[width=5.5cm,height=5.5cm]{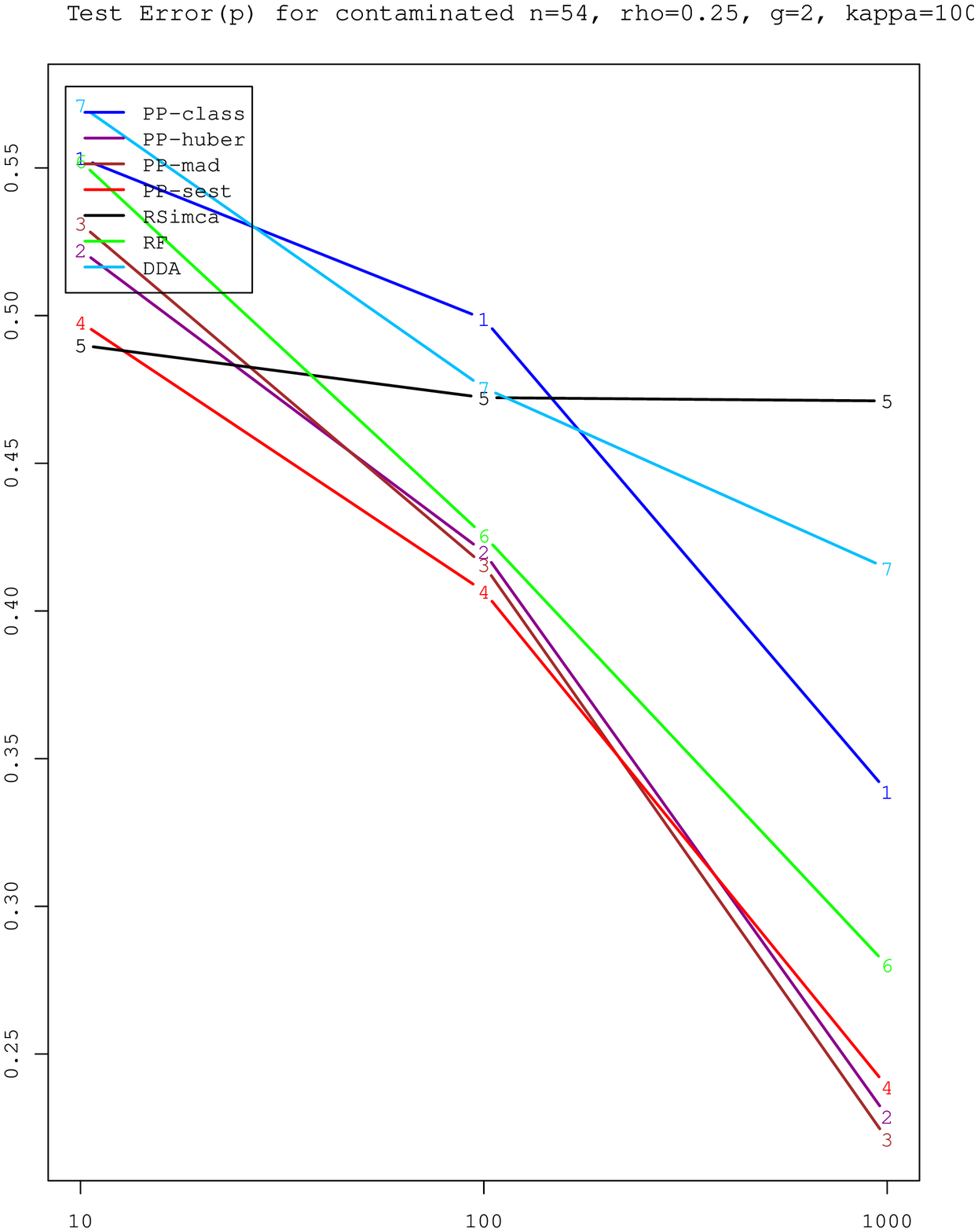}
  \caption{Average test error on the strongly contaminated simulated  data with $g=2$ and
  $\rho=0.25$. We herein reveal for each of the 7 methods, the effect of the input space dimension $p$ on the average test error.}
  \label{fig:sim:strong:pp:2}
\end{figure}

\begin{figure}[!htbp]
  \centering
  \includegraphics[width=5.5cm,height=5.5cm]{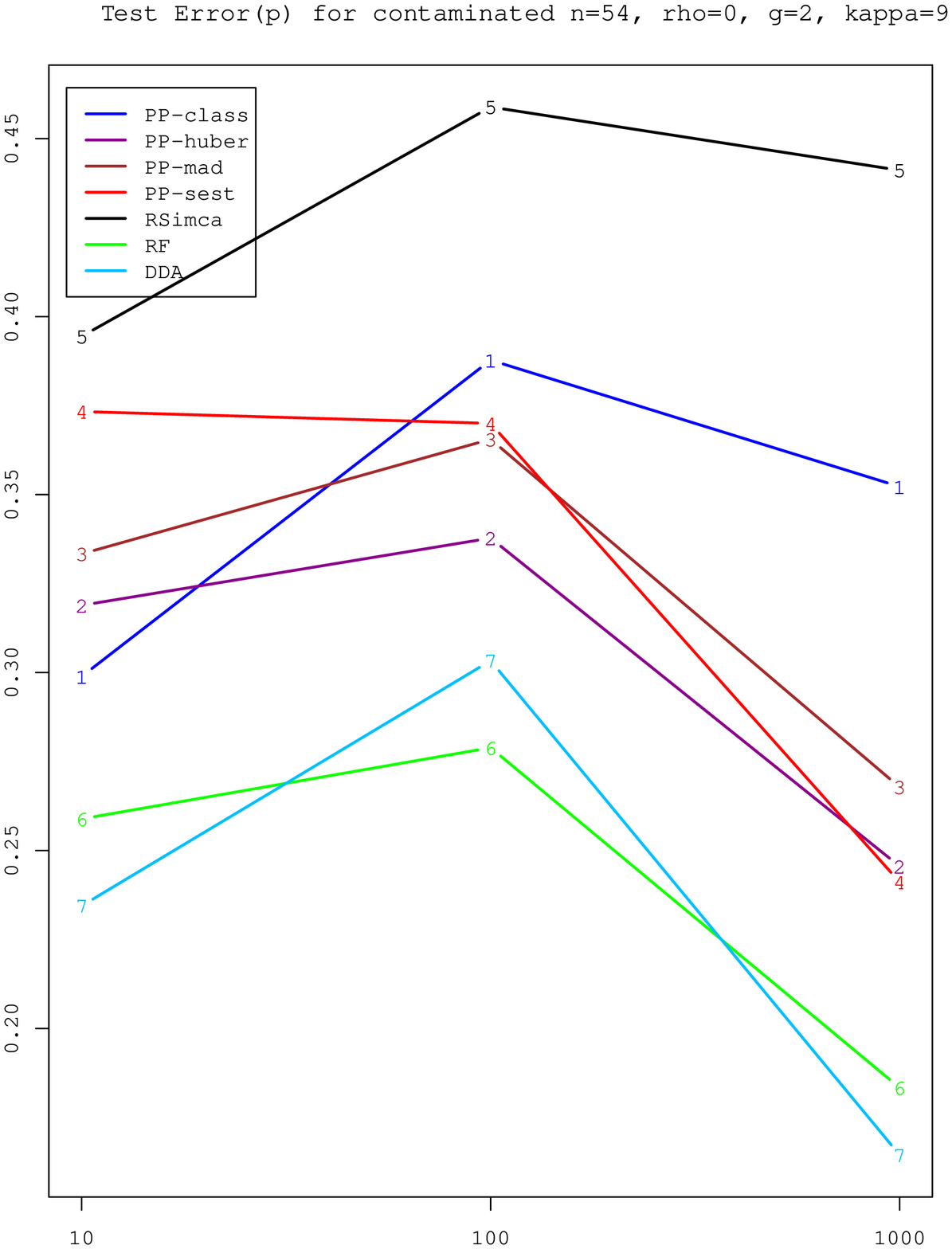}
  \includegraphics[width=5.5cm,height=5.5cm]{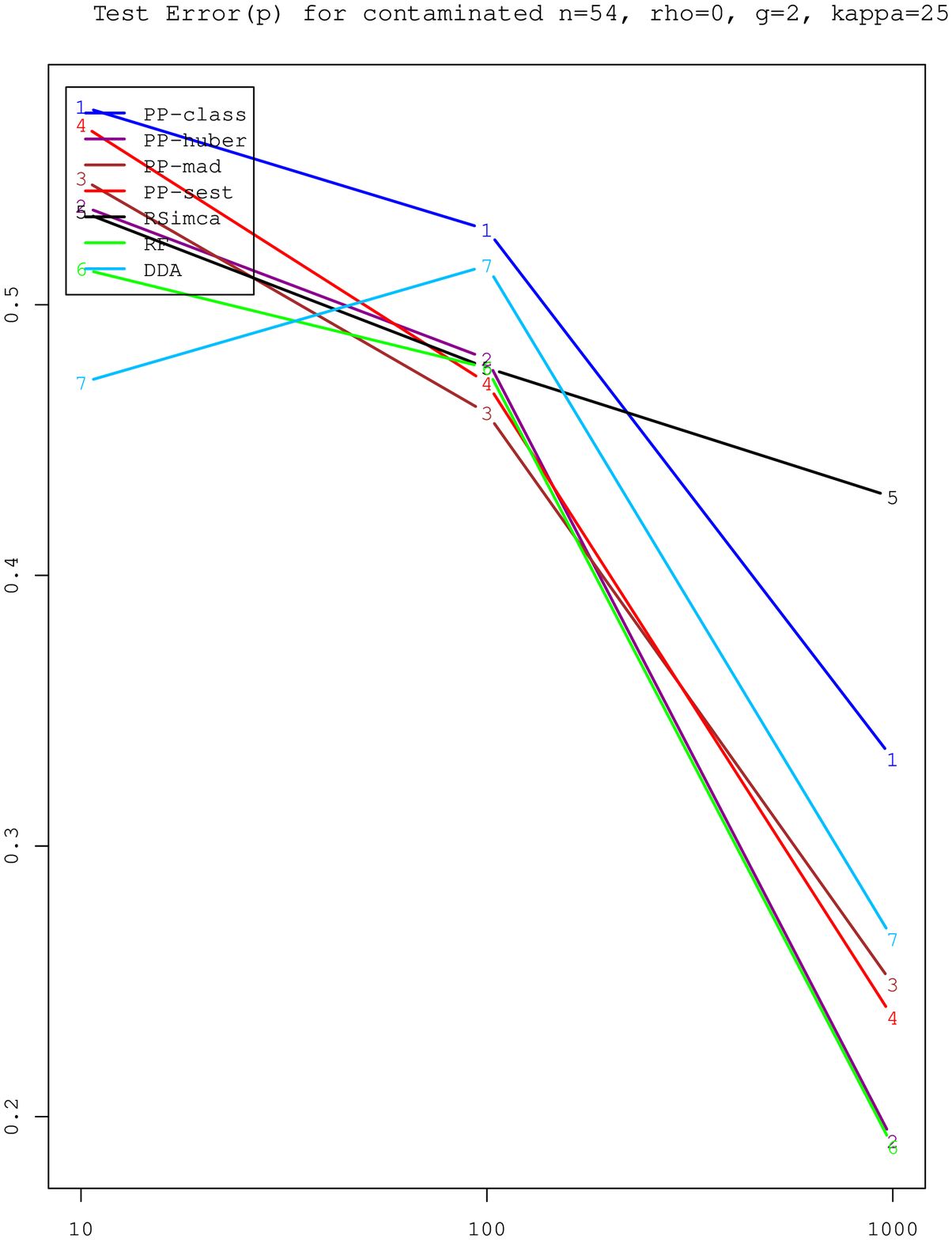}
  \includegraphics[width=5.5cm,height=5.5cm]{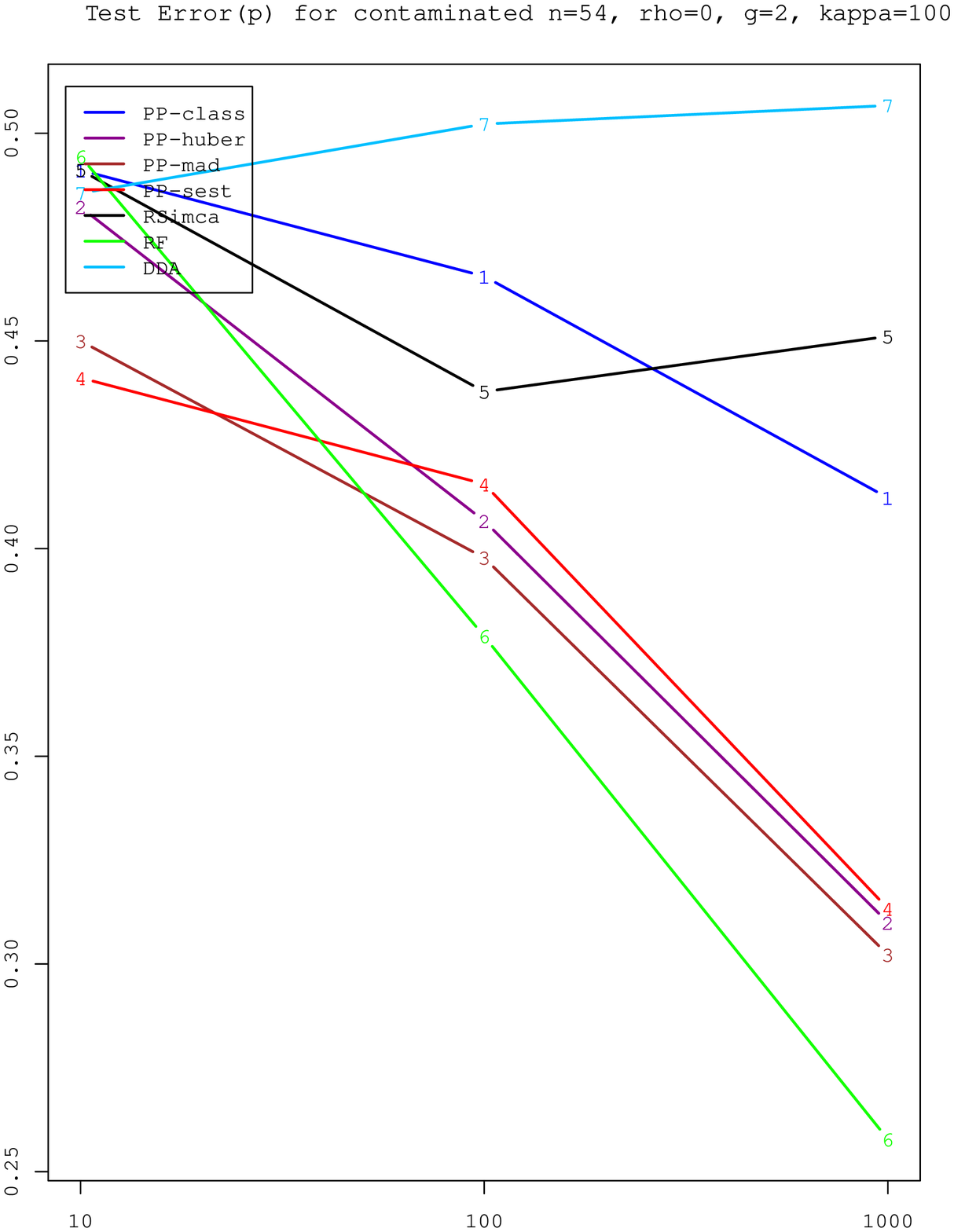}
  \caption{Average test error on the strongly contaminated simulated  data with $g=2$ and
  $\rho=0.00$. We herein reveal for each of the 7 methods, the effect of the input space dimension $p$ on the average test error.}
  \label{fig:sim:strong:pp:3}
\end{figure}

\begin{figure}[!htbp]
  \centering
  \includegraphics[width=5.5cm,height=5.5cm]{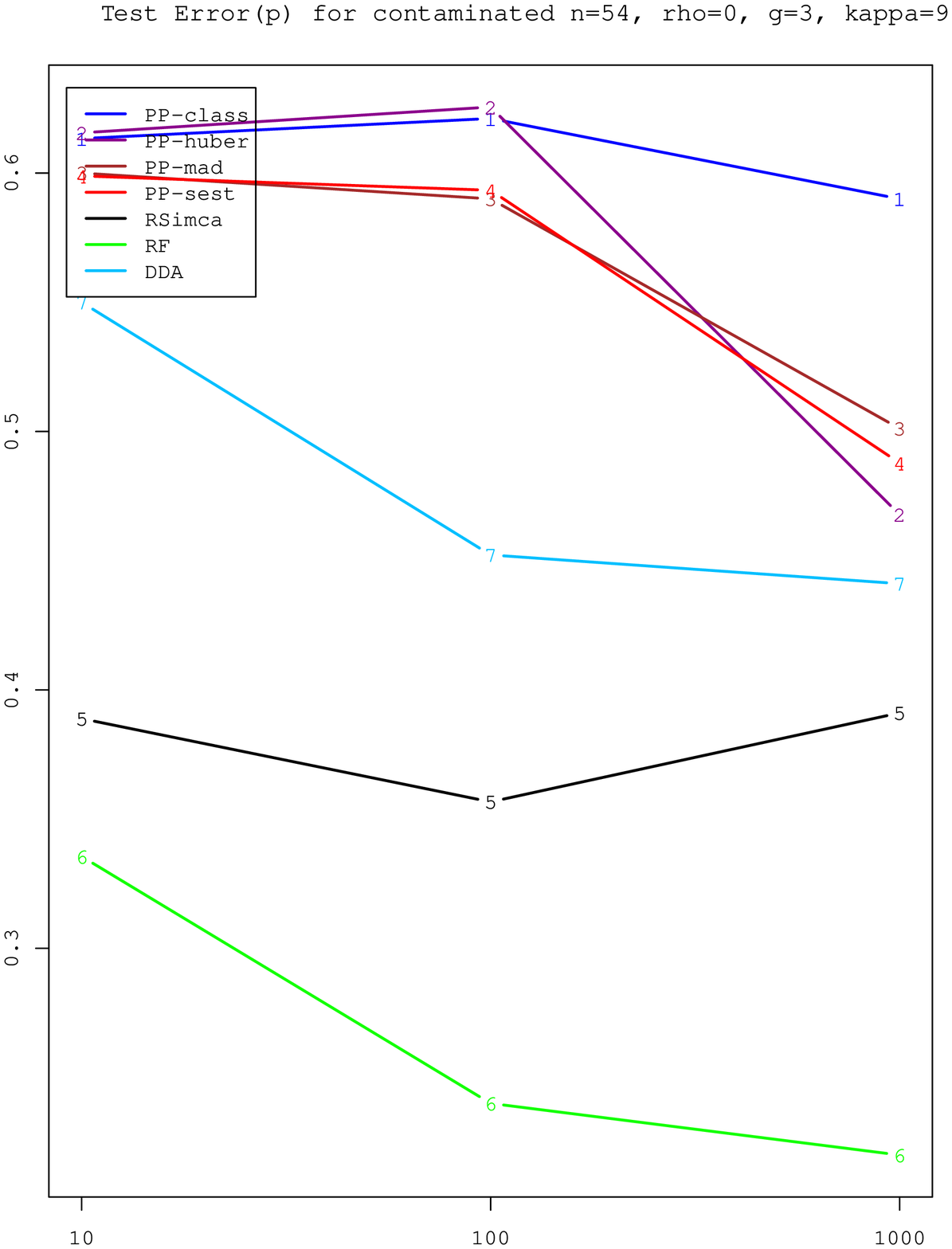}
  \includegraphics[width=5.5cm,height=5.5cm]{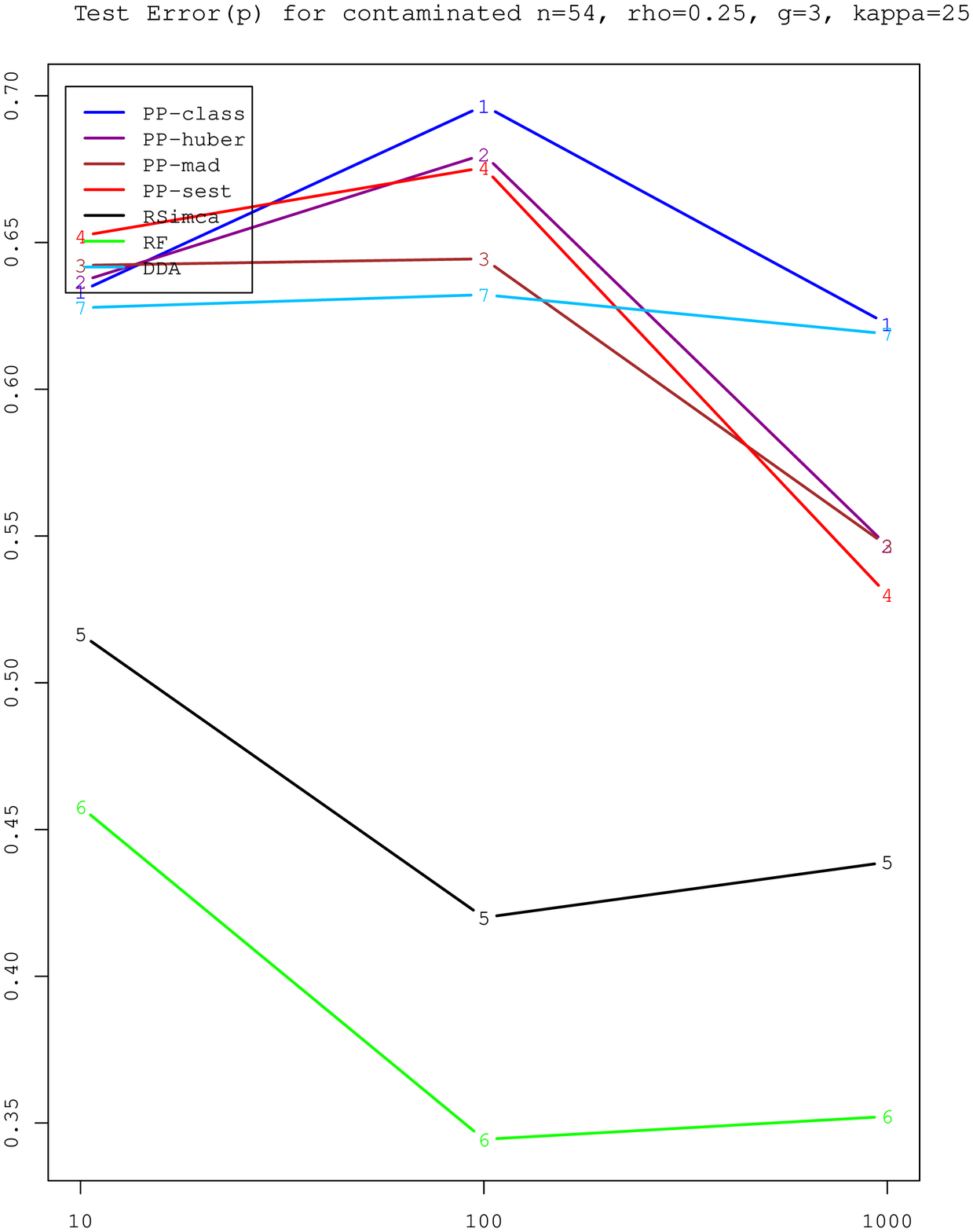}
    \includegraphics[width=5.5cm,height=5.5cm]{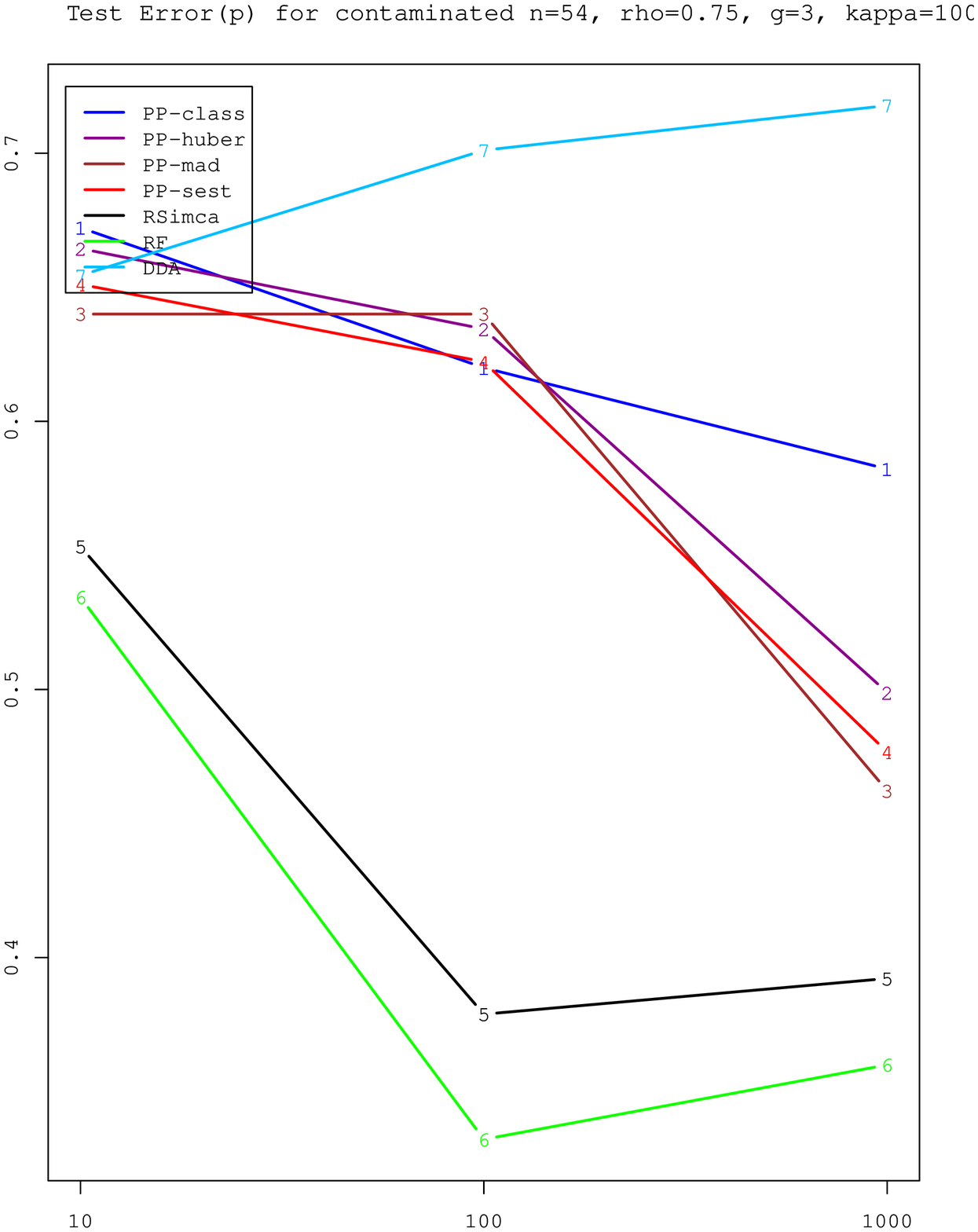}
  \caption{Average test error on the strong contaminated simulated  data with $g=3$.
  We herein reveal for each of the 7 methods, the effect of the input space dimension $p$ on the average test error.}
  \label{fig:sim:strong:pp:4}
\end{figure}

We see in Figure \eqref{fig:sim:strong:pp:4} that under the strong contamination regime,
RF emerges as the best in multi-categorical classification tasks, regardless of the correlation level
and the size $\kappa$ of contamination. SIMCA also performs very well under these conditions,
typically taking the second place to RF in predictive performance.
It can said that overall, RF and SIMCA are the most practical and applicable of all the techniques explored here,
since they both do well for the realistic scenario of multicategorical classification with some rate of contamination.

\section{Conclusion and Discussion}
We have presented a thorough comparison of the predictive performances of several robust classification methods
on high dimension low sample size data.  On both real life and simulated data, interesting patterns emerged.
We noted for instance that the SIMCA method, by being somewhat very general tends to yield mediocre predictive performances
when $p$ is much larger than $n$, even though it rarely yield the worst among compared classification techniques.
One of the most striking remarks in our study has to do with projection pursuit, the clearest being the fact that
projection pursuit seems to do well only in binary classification. As a matter of fact, for all the scenarios involving
more than two classes, projection pursuit seems to fail miserably regardless of all the other aspects of the data.
Strikingly also, when there are only two classes, projection pursuit yields the best predictive performance if
the correlation among the input space variables is large. This leads us to conclude that projection pursuit as
a method for robust discriminant analysis is - at least in its present form - only best suited to  binary classification
for data whose intrinsic dimensionality is very low. For us, the most striking observation lies with the performance
of random forest. Indeed, as can be noted in all the computational results presented earlier, random forest
tended to be the  best overall. More precisely, there was no instance where random forest yielded the worst performance,
and in most cases, it was either the very best or the second best. As we explained earlier, this can be explained
by the very mechanism of random forest in the sense that at every iteration of the construction of a random forest,
not only is the estimator based on a subset of input variables, but also crucially the bootstrap mechanism leaves
out a proportion $e^{-1}$ of the sample. This left out fraction certainly contains some of the outliers.
It is our conjecture as indicated earlier, that the fact of leaving out a fraction of the data
allows random forest to weed out outliers or at least average out their effect. Hence the inherent ability of
random forest to achieve robustness by random subsampling.
One could conjecture that the overall superior performance of random forest can be attributed to the fact
it does both variable selection (by random subspace learning) thereby inherently addressing the extremely high dimensionality of the data,
and also reduction (or even elimination) of the effect of outliers by subsampling.
There is a sense here of a connection - however loose - between
the subset selection of minimum covariance determinant (MCD) - recall that MCD select $h<n$ observations that yield the minimum
covariance determinant - and the out-of-bag observations derived from the bootstrap in random forest. Therefore,
we intend to investigate further the relationship between the fraction/proportion $e^{-1}$ of random forest
and the number $h$ of observations used the MCD. We are also currently exploration various
strategies of regularized MCD as a way to achieve robust classification in settings where $n$ is much less than $p$.

\section*{Acknowledgements}
Ernest Fokou\'e wishes to express his heartfelt gratitude and infinite thanks to Our Lady of Perpetual Help for Her
ever-present support and guidance, especially for the uninterrupted flow of inspiration received through Her
most powerful intercession.

\bibliographystyle{chicago}
\bibliography{gf-robust-ref}
\end{document}